\documentclass[oneside, twocolumn, 10pt, a4paper]{article}
\usepackage[left=1.5cm, right=1.5cm, top=2.5cm, bottom=2.5cm]{geometry}
\usepackage[utf8]{inputenc}
\usepackage[english]{babel}
\usepackage{authblk}
\usepackage[version=3]{mhchem}
\usepackage{csquotes}
\usepackage{layouts}
\usepackage{abstract}
\usepackage{graphicx} 
\usepackage{amsmath}
\usepackage{braket}
\usepackage{balance}
\usepackage{siunitx}
\DeclareSIUnit\molpercent{mol\%}
\usepackage{amssymb}
\usepackage{amsmath}
\usepackage{enumitem}
\usepackage{placeins}
\usepackage{float}
\newcommand{\abs}[1]{\left \lvert #1 \right \rvert} 
\newcommand{\dev}{\text{d}} 

\usepackage[style=chem-rsc, doi=true]{biblatex}
\usepackage{hyperref}

\title{Spectroscopic Analysis of Vibrational Coupling in Multi-Molecular Excited States}
\author[1,4]{Sebastian Hammer\thanks{mail to: sebastian.hammer@mail.mcgill.ca}}
\author[2]{Theresa Linderl}
\author[1]{Kristofer Tvingstedt}
\author[2]{Wolfgang Brütting}
\author[1,3]{Jens Pflaum\thanks{mail to: jens.pflaum@physik.uni-wuerzburg.de}}

\affil[1]{Experimental Physics VI, Julius Maximilians University Würzburg, 97074 Würzburg, Germany.}
\affil[2]{Institute of Physics, University of Augsburg, 86135 Augsburg, Germany.}
\affil[3]{Bavarian Center for Applied Energy Research, 97074 Würzburg, Germany.}
\affil[4]{Present address: Center for the Physics of Materials, Department of Physics and Department of Chemistry, McGill University, Montreal, QC H3A 2K6, Canada.}

\date{}

\begin{document}

\twocolumn[
\maketitle
    \begin{onecolabstract}
        Multi-molecular excited states accompanied by an intra- and inter-molecular geometric relaxation are commonly encountered in optical and electrooptical studies and applications of organic semiconductors as, for example  excimers or charge transfer states. Understanding the dynamics of these states is crucial to improve organic devices such as light emitting diodes and solar cells. Their full microscopic description, however, demands for sophisticated tools such as ab-initio quantum chemical calculations which come at the expenses of high computational costs and are prone to errors by assumptions as well as iterative algorithmic procedures. Hence, the analysis of spectroscopic data is often conducted on a phenomenological level only. Here, we present a toolkit to analyze temperature dependent luminescence data and gain first insights into the relevant microscopic parameters of the molecular system at hand. By means of a Franck-Condon based approach considering a single effective inter-molecular vibrational mode and different potentials for the ground and excited state we are able to explain the luminescence spectra of such multi-molecular states. We demonstrate that by applying certain reasonable simplifications the luminescence of charge transfer states as well as excimers can be satisfactorily reproduced for temperatures ranging from cryogenics to above room temperature. We present a semi-classical and a quantum-mechanical description of our model and, for both cases, demonstrate its applicability by analyzing the temperature depended luminescence of the amorphous donor-acceptor heterojunction tetraphenyldibenzoperiflanthene:\ce{C60} as well as polycrystalline zinc-phthalocyanine to reproduce the luminescence spectra and extract relevant system parameters such as the excimer binding energy.    
    \end{onecolabstract}
    ]
    \saythanks

\clearpage
\section{Introduction}
The photo-physics of molecular semiconductors is usually described by excitons, i.e. partially delocalized electronically excited molecular states, and their interaction with neighboring entities which is summarized by Kasha's exciton theory   \cite{Kasha_1965}.  As the electronic coupling of adjacent molecules depends on their relative orientation, the underlying crystal structure strongly affects the photo-physics and its dynamics in molecular aggregates  \cite{Gierschner_2013, Spano2016, Gieseking2014}, one prominent example being the occurrence of Davydov-splitting  \cite{Davydov1948, Davydov_1964}.\par
However, as it has been shown in recent studies on more complex compounds, Kasha theory only provides a simplistic picture on the underlying physical mechanisms \cite{Hestand_2018, Engels_2017} as it, for instance, ignores molecular vibrations as well as deformation of the crystalline lattice and its molecular constituents upon excitation. To overcome these shortcomings, one approach is to only consider the smallest possible unit of the system at hand that is necessary to describe its photo-physics and to calculate all adiabatic nuclear and electronic relaxations involved on a full quantum-mechanical level  \cite{Fink_2008, Engels_2017}. The main disadvantage of this approach, apart from the huge computational effort, is the inherent confinement of the excitation to a few molecules as the interaction with more distant molecules is only considered as a  mean dielectric screening of the central molecular group. If the crystallinity of the system and its intra-molecular vibrations are taken into account, the excitation can be described in terms of an exciton-polaron, i.e. a photo-excited state within a single crystal, which is dressed by intra-molecular vibrations extending to adjacent molecules  \cite{Spano_2010}. Formally, the description is given by a Holstein-Peierls model which takes into account both the vibrational energy and the reorganization due to vibronic coupling of the excited state  \cite{Hestand_2018}. This approach, however, treats the crystal's molecular constituents as fixed in position and, hence, is not suited to describe states originating from inter-molecular reorganization. Only recently Bialas and Spano successfully extended the Holstein-Peierls approach to describe inter-molecular geometry relaxation and the subsequent radiative relaxation  \cite{Bialas2022}. \par
Emission from excited states which emerge from an inter-molecular geometry relaxation, e.g. excimers or exciplexes (vide infra), are common in molecular single crystals as well as polycrystalline thin films \cite{Ferguson1958, Tanaka1963, Gierschner_2013, Hausch2022} and are often found in organic light emitting diodes (OLEDs)  \cite{Kalinowski2009, Uoyama2012, Sarma2018, Bunzmann2020} and solar cells  \cite{Deibel2010, Vandewal2016}. Usually, the emission from such states is characterized by a broad and unstructured spectrum which is considerably red shifted compared to the single molecule emission  \cite{Birks1964,Foerster1969,Gierschner_2013}. A unique feature of such excited multi-molecular states is the temperature dependence of their emission spectrum showing a characteristic broadening with increasing temperature  \cite{Birks1968, Tvingstedt_2020, Bialas2022, Hong2022}\par
For donor-acceptor heterojunctions in solar cells it is common since many years to describe the absorption and emission spectra of charge-transfer (CT) states in terms of Marcus theory for which the inter- and intra-molecular geometrical relaxation is mediated by low energy vibrations. Two main concepts have emerged: In the picture of \textit{dynamic disorder} the broadening of the spectra is solely the product of inter- and intra-molecular vibrations  \cite{Vandewal2010, Vandewal2010a,  Tvingstedt_2020, Benduhn2017, Goehler2021}. The \textit{static and dynamic disorder} based modelling, in contrast, includes an energetic disorder term  in the range of \SIrange{50}{100}{\meV} to account for structural inhomogenities. This is the main reason for the broad spectra recorded at low temperatures and the finite line width found for the extrapolation to $T \rightarrow \SI{0}{\kelvin}$ while low energy vibronic transitions only influence the spectra at higher temperatures  \cite{Burke2015, Kahle2018, Linderl_2020, Melianas2019, Yan2021}. Both pictures describe the experimental observations in various material systems well and, as a consequence, no general consensus has emerged, so far. \par
Here, we present a model that explains the temperature dependent evolution of the luminescence spectra of multi-molecular excited states solely by their relaxation within a modified potential energy surface (PES) after excitation and the subsequently induced inter-molecular vibrations. We introduce a semi-classical as well as a quantum-mechanical description based on a displaced harmonic oscillator model which, in difference to other approaches, is based on unequal ground and excited state potentials. We  show that the evolution of the luminescence spectra with temperature is sufficient to qualitatively asses the potential energy landscape along the geometric relaxation coordinate. Finally, our model is validated for a prototypical charge-transfer system, comprising tetraphenyldibenzoperiflanthene (DBP) and fullerene \ce{C60}  \cite{Linderl_2020}, as well as for the excimer system zinc-phthalocyanine (ZnPc). We demonstrate that the proposed model is indeed able to reproduce all spectral key features of the emission over a broad temperature range and, thus, yields access on the essential parameters characterizing the system's PES, such as the vibrational energy quantum of the inter-molecular modes as well as binding and relaxation energies.

\section{Modeling multi-molecular emission}
\begin{figure}[htb]
    \centering
    \includegraphics{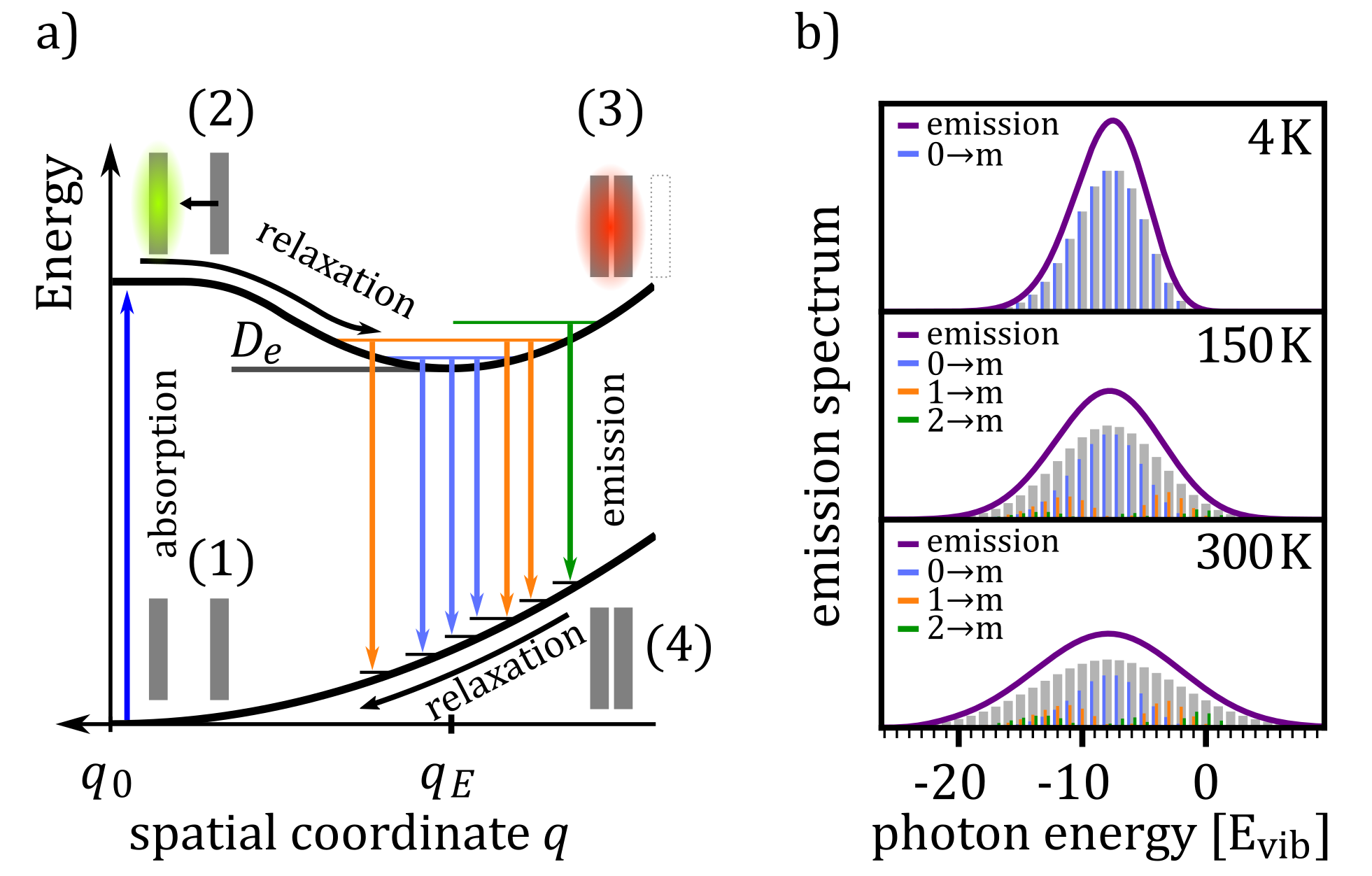}
    \caption{a) Schematic representation of X-dimer formation and emission: After photo-excitation of a dimer (1) an exciton-polaron is formed (2) which adiabatically relaxes to an energetically favorable ground state via changing the dimer geometry (3). Thermally populated inter-molecular vibrations couple to the electronically excited state leading to a Franck-Condon type emission to higher lying vibrational levels of the dimer ground state (4). From there, the dimer relaxes back to its ground state geometry (1). b) X-dimer emission spectra for three different temperatures calculated in a displaced harmonic oscillator model. The photon energy is given in units of the vibrational energy quantum relative to the 0-0 transition marked by 0. The stick spectra (grey) indicate the intensity of the vibronic transitions. The individual contributions from the X-dimer's ground, first and second vibrationally excited state to the emission are shown in blue, orange and green, respectively.}
    \label{fig:fig1}
\end{figure}

Figure \ref{fig:fig1} a) shows the excitation scheme of a simple bi-molecular excitation within a geometric relaxation model: Two molecules located on adjacent lattice sites at equilibrium distance $q_0$ mark the starting point of the excitation-relaxation cycle (1). Via photo-excitation one molecule is elevated to an electronically excited state, yielding a Frenkel exciton-polaron (2). This excited state now evolves, accompanied by an adiabatic relaxation of the inter-molecular geometry, to a lower lying state, for which (i) the inter-molecular geometry is different from that of the ground state equilibrium position and (ii) the excited state's wavefunction is delocalized over the participating adjacent molecules. For the case shown in figure \ref{fig:fig1} a), the inter-molecular distance between the two molecules is reduced from $q_0$ to $q_E$ and the excitation delocalizes equally over both monomers (3). For two identical molecules this state is commonly referred to as an \textit{excimer}, a portmanteau of \textit{excited} and \textit{dimer}. In case of a dimer comprising a donor and acceptor molecule, this results either in a charge-transfer (CT) complex or an exciplex (\textit{excited complex}), depending on the ground state interaction  \cite{Belova_2017}. For the sake of generality, we will call such a state \textit{X-dimer} from here on, deliberately leaving the underlying resonance interaction unspecified. The X-dimer entity, bound by its electronic potential, is subject to inter-molecular vibrations. Due to the different geometry compared to the surrounding crystal, these vibrations are confined to the dimer entity and, in a first approximation, are decoupled from the crystal phonons. The X-dimer's vibrational states become thermally occupied forming a vibrationally and electronically excited dimer state which, finally, can decay radiatively to the ground state. In the diabatic Franck-Condon picture, the electronic transitions are much faster than the attending changes in the inter-molecular geometry which leads to the generation of a \textit{hot ground state}, i.e. an electronic ground state dressed by high energy phonons (4). The crystal relaxes back to its equilibrium position  via thermalization. \par
From the description above, it becomes clear that some simplifications have been made in the description of the formation and relaxation processes. First of all, the excited state interaction leading to the geometric relaxation has not been specified in detail, which actually is not necessary for the phenomenological description of the X-dimer's photon emission. In the case of excimer formation, exciton resonance \cite{Foerster1962, Foerster1963} as well as a CT state  \cite{Ferguson1958} have been proposed as driving forces behind geometrical relaxations but, meanwhile, the consensus prevailed that only an adiabatic mixing of Frenkel and CT excitons can fully explain the observed phenomena  \cite{Azumi1964, Azumi1964a, Kalinowski2009, Casanova2015, Ramirez2020, Hong2022, Bialas2022}. For example, in crystalline pentacene, the first excited state is known to have a significant admixture of CT character \cite{Hestand2015, Craciunescu2022} and a considerable geometric reorganization accompanies its photo-excitation \cite{Seiler_2021}. In fact, Tvingstedt et al. successfully used a displaced harmonic oscillator model using inter- and intra-molecular vibrational modes to describe the emission of CT states formed at donor-acceptor hetero-junctions \cite{Tvingstedt_2020}. The evolution from the inital excitation towards the excimer state is often considered a single-step process \cite{Foerster1969, Birks1975, Hoche_2017, Hoche2021}, but a two step process via an intermediate state has been proposed, too \cite{Cook2017, Pensack2018}, which can be part of the adiabatic evolution towards the relaxed excimer geometry  \cite{Hong2022}. Furthermore, the geometric reorganization upon photo-excitation usually comprises more than just one motion  \cite{Gierschner2003, Casanova2015}. For example, calculations revealed the pyrene excimer formation to be governed by a superposition of a lateral convergence as shown in  figure \ref{fig:fig1} a) and a horizontal sliding motion  \cite{Hoche_2017}. For perlyene bisimide (PBI) dimers, a relative rotation of the stacked PBI dimer is predicted to govern the geometric relaxation upon photo-excitation \cite{Fink_2008, Hong2022}. This means, the relaxation path shown in figure \ref{fig:fig1} a) between configurations (2) and (3) should be interpreted as an energetially steered pathway along which the system relaxes towards a local minimum of its PES and, hence, $q$ can be interpreted as a generalized coordinate. This implies, that the inter-molecular vibrations of the relaxed dimer state in configuration (3) are composed of several vibrational modes, e.g. shifting and sliding motions \cite{Warshel_1974, Gierschner2003, Hoche_2017}. We will show that for many applications, the assumption of one dominant vibrational mode governing the vibronic transitions, as it is commonly applied for intra-molecular vibrations in molecular spectroscopy, can be a useful strategy to satisfactorily explain X-dimer emission spectra and to obtain first insights into the system's PES and relaxation pathways. \par
For the case of a single vibrational mode coupling to the X-dimer and its electronic ground state, the emission process is a transition from an electronically excited dimer state $\Ket{X,n}$ at vibrational state $n$, to the electronic ground state $\Ket{G,m}$ in its vibrationally excited state $m$. The probability of thermal population of the vibrationally excited state $\Ket{X,n}$ at temperature $T$ is given by the Boltzmann probability $P(n,T)=Z^{-1}\exp\left(-(n+\frac{1}{2})E_\text{X,vib}/k_\text{b}T\right)$ with the canonical partition function $Z=\sum_z \exp\left(-(z+\frac{1}{2})E_\text{X,vib}/k_\text{b}T\right)$ and $E_\text{X,vib}$ as the vibrational energy quantum of the X-dimer potential. In a Franck-Condon picture, the emission spectrum \cite{note_emission_spectrum} is then given by

\begin{equation}
    \resizebox{0.85\columnwidth}{!}{$\displaystyle
    \overline{I} \left( E, T \right) \dev E= \sum_n \sum_m P(n,T) \abs{\Braket{m|n}}^2 \Gamma \left(E- \Delta E_{nm}, \sigma \right)$.}
    \label{eq1}
\end{equation}

Here, $\Gamma \left(E- \Delta E_{nm}, \sigma \right)$ is a line shape function with a line width parameter $\sigma$ and $\Delta E_{mn}$ is the photon energy of the radiative transition$\Ket{X,n} \rightarrow \Ket{G,m}$. The Franck-Condon factor $\abs{\Braket{m|n}}^2$ is defined by the overlap integral of the nuclear wavefunctions $\Ket{n}$ and $\Ket{m}$. If ground state and X-dimer potential are of the same form, \textit{i.e.} both resemble that of a displaced harmonic oscillator model, Thomas Keil derived a general expression for the Franck-Condon factor \cite{Keil_1965} given by 
\begin{equation}
\resizebox{0.85\columnwidth}{!}{$\displaystyle
\abs{\Braket{m|n}}^2 = 
	\begin{cases}
	e^{-S} S^{n-m} \left( \frac{m!}{n!} \right) \left( L_{m}^{n-m}(S) \right)^2, & \text{for } n \geq m \\
	e^{-S} S^{m-n} \left( \frac{n!}{m!} \right) \left( L_{n}^{m-n}(S) \right)^2, &\text{for }n \leq m					
	\end{cases}$.}
	\label{eq2}
\end{equation}
where $ L_{m}^{n-m}$ are the associated Laguerre polynomials and $S$ is the Huang-Rhys parameter  \cite{de_Jong_2015}. \par
Figure \ref{fig:fig1} b) shows the emission line shape of an X-dimer at three different temperatures calculated from equation \eqref{eq1} utilizing the displaced harmonic oscillator approximation \eqref{eq2} with a vibrational energy quantum of \SI{12}{\meV} and a Huang-Rhys parameter of $S=8$. The calculated stick spectra displayed in grey indicate the intensity at steps of the vibrational energy with 0 marking the energy of the vibronic 0-0 transition on the energy axis. The contribution of the individual vibrational levels of the X-dimer are color coded. The superposition of gaussian lineshape functions with a line width of \SI{10}{\meV} for each vibronic emission generates the broad envelope. Evidently, at \SI{4}{\kelvin} the emission is dominated by transitions from the X-dimer's vibrational ground state and the large vibronic coupling ($S=8$) causes a preferential transition to high vibrationally excited levels of the ground state resulting in an asymmetric emission line shape with a pronounced low energy flank. 
With increasing temperature, higher vibrational levels of the X-dimer are populated and contribute to the overall emission yielding a redistribution of emission intensity between the different vibronic transitions. This leads to a broadening of the emission spectrum while simultaneously the intensity maximum decreases. Although the displaced harmonic oscillator model can be used to describe the emission of CT states  \cite{Tvingstedt_2020, Benduhn2017}, the assumption of the same inter-molecular potential for the electronic ground and excited state turns out to be justified only at first approximation taking into account the large geometric reorganizations expected upon X-dimer formation. Therefore, in the following we present an extended approach to this problem based on different ground and excited state potentials treated in a semi-classical as well as a quantum mechanical description. 

\subsection{Semi-classical description}
The semi-classical description treats the X-dimer's inter-molecular vibrations as a quantum mechanical harmonic oscillator while classically approximating the ground state potential to be continuous as suggested by Birks and Kazzaz  \cite{Birks1968}. This approximation holds true under two conditions: 
\begin{enumerate}[label=(\roman*)]
    \item \label{it.cond1} The displacement of the oscillators is large in relation to the vibrational energy quantum of the final state  \cite{Note_condition1_S-large} leading to a population of vibrational states with large quantum numbers  \cite{Williams_1951, Lax_1952}. 
    \item \label{it.cond2} The vibrational energy quantum of the final state, here the ground state, is smaller or of the same order of magnitude as the line width parameter, i.e. $E_\text{G,vib} \lesssim 2\sigma$. 
\end{enumerate}
The final states described by condition \ref{it.cond1} are characterized by wave functions whose largest amplitude and, thus, contribution is located at the classical turning points of the oscillator \cite{Williams_1951, Lax_1952}, concentrating the transition probability described by the Franck-Condon factor in equation \eqref{eq1} to a small region of the nuclear displacement \cite{note_injective_relation} $\Delta q \rightarrow 0$. Condition \ref{it.cond2} ensures that the individual vibronic transitions are close in energy justifying a continuous ground state potential $R(q)$. \par

\begin{figure*}[!htb]
    \centering
    \includegraphics{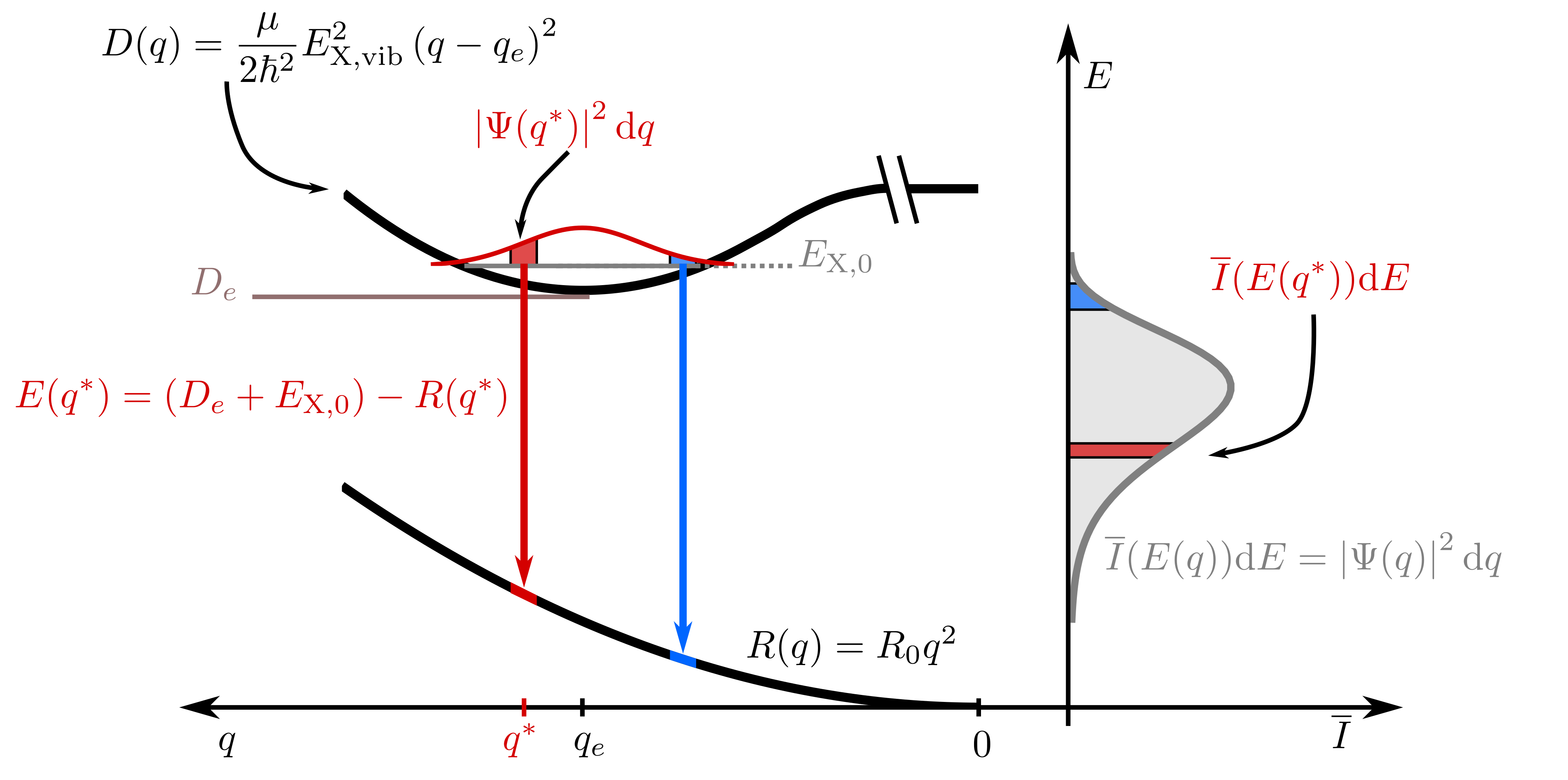}
    \caption{Semi-classical approach to calculate the X-dimer emission spectra. The q abscissa of the inter-molecular distance points, by definition, in positive direction from right to left by definition. The excited state is assumed to behave like a harmonic oscillator at energy offset $D_e$ radiatively relaxing into a quasi continuous ground state of repulsive potential $R(q)$. The spectral density as function of the photon energy $\overline{I}(E)\dev E$ is then directly related to the spatial probability distribution of the harmonic oscillator given by the squared absolute of the vibrational dimer wave function $\abs{\Psi(q)}^2$.}
    \label{fig:fig2}
\end{figure*}

Without loss of generality, we now define the sign of the general nuclear displacement coordinate $q$ to be positive along the path on the PES towards the new energetic minimum after photo-excitation which defines the X-dimer's fully relaxed geometry. Furthermore, the equilibrium position in the electronic ground state is set to $q_0 := 0$, as shown in figure \ref{fig:fig2}. Here, the excimer equilibrium position $q_e$ has a positive value and in the picture of a lateral convergence described in figure \ref{fig:fig1} a), $q_e = \abs{q_E-q_0}$ is the absolute value of molecular displacement of both entities forming the X-dimer with respect to the ground state equilibrium $q_0$. In the semi-classical picture, the ground state potential is given as
\begin{equation}
    R(q) = R_0 q^2 = \frac{\mu}{2\hbar^2}E_\text{G,vib}^2 q^2
    \label{eq3}
\end{equation}
with the oscillator constant $R_0 = \frac{\mu}{2\hbar^2}E_\text{G,vib}^2$ expressed by the reduced mass \cite{Note_mass_inertia} $\mu$ and the vibrational energy quantum of the ground state oscillator. In the same manner the X-dimer potential can be expressed as 
\begin{equation}
    D(q)=\frac{\mu}{2 \hbar^2}E_\text{X,vib}^2 \left(q-q_e \right)^2 + D_e 
    \label{eq4}
\end{equation}
defining the quantum mechanical oscillator with the vibrational energy quantum $E_\text{X,vib}$ and the energetic offset $D_e$ with respect to the ground state minimum set at $E=0$. The total energy of the X-dimer state $\Ket{X,n}$ is then given by 
\begin{equation}
    E_\text{X,n} = D_e + \left(n+\frac{1}{2} \right) E_\text{X,vib}.
    \label{eq5}
\end{equation}
A photon emitted by a radiative transition from $\Ket{X,n}$ to a final state with displacement configuration $q^*$ has the energy
\begin{equation}
    E(q^*) =  E_\text{X,n} - R(q^*).
    \label{eq6}
\end{equation}
Two exemplary transitions are shown in figure \ref{fig:fig2} (red and blue arrow) for the X-dimer's vibrational ground state $n=0$ and two different final state geometries. The probability of the X-dimer geometry to adopt the displacement $q^*$ is given by the probability density $\abs{\Psi(q^*)}^2\dev q$ with $\Psi(q)$ being the nuclear wave function of the X-dimer state. Hence, the emission spectrum $\overline{I}(E)\dev E$ for a photon energy $E$ is related to the spatial probability distribution by 
\begin{equation}
    \overline{I} \left( E \left( q^* \right) \right) \dev E = \abs{ \Psi \left(q^* \right) }^2\dev q.
    \label{eq7}
\end{equation}
This is illustrated in figure \ref{fig:fig2} by the red and blue intervals for lower and higher emission energies, respectively. The overall emission spectrum is given by the grey curve in the right part of figure \ref{fig:fig2} with the energy axis as its vertical abscissa. From equation \eqref{eq6} the spatial coordinate $q$ can be expressed as an inverse function of the photon energy $\tilde{q}_n(E)$ for each vibrationally excited state $\Ket{X,n}$. Assuming harmonic oscillator wave functions $\Psi_n$ for the X-dimer, the emission from the vibrational ground state $\Ket{X,0} \rightarrow \Ket{G}$ plotted as the grey spectrum in figure \ref{fig:fig2} is given by
\begin{equation}
    \overline{I}_0\left( E \right) \dev E = \frac{1}{2} \sqrt{\frac{\alpha}{R_0 \pi \left( E_\text{X,0}-E \right)}}\exp \left(- \alpha \Tilde{q}_0^2 \right) \dev E
    \label{eq8}
\end{equation}
with $\alpha = \mu E_\text{X,vib}/ \hbar^2$. For mathematical details and the respective analytical expression for the five lowest vibrationally excited states we refer to supplementary note 1 in the supplementary information (SI).
Using the same approach for each vibrationally weighted spectrum of the transitions $\Ket{X,n} \rightarrow \Ket{G}$ the temperature dependent emission given by \eqref{eq1} reduces to
\begin{equation}
        \overline{I} \left(E, T \right) \dev E = \sum_n  P \left(n,T \right) \overline{I}_n\left( E \right) \dev E.
        \label{eq9}
\end{equation}\par
As it becomes evident by equation \eqref{eq8} the analytical form of the emission spectrum is only valid for energies $E<E_\text{X,0}$ due to the singularity at $E=E_\text{X,0}$ and the consecutive negative values in the square-root term rendering any values of $ \overline{I} \left(E, T \right) \dev E$ above $E_\text{X,0}$ to be complex and thus unphysical. Moreover, this statement can be generalized for higher vibrational levels, constricting the valid energy range of each emission line shape $ \overline{I}_n(E)$ to $E<E_\text{X,n}$. While for a strong X-dimer binding energy and low quantum numbers $n$ the expression of the overall emission is not impeded, at higher temperatures not all contributions to the sum in equation \eqref{eq9} can be evaluated at higher emission energies. This leads to practical limitations in the high temperature limit. Hence, in the next section, this shortcoming will be overcome by extending the model to a full quantum mechanical model. 

\subsection{Quantum mechanical description}
The quantum mechanical description of the X-dimer emission spectrum is given by the general expression in equation \eqref{eq1}. Thus, for two displaced harmonic oscillators with distinct potentials, the Franck-Condon factor $\abs{\Braket{m|n}}^2$, defining the intensity for a transition $\Ket{X,n} \rightarrow \Ket{G,m}$, needs to be evaluated. For a ground state and an excited state described by the potentials \eqref{eq3} and \eqref{eq4} with the harmonic oscillator wave functions $\Phi_m$ and $\Psi_n$, respectively, the overlap integral of the Franck-Condon factor is formally expressed by
\begin{equation}
    \Braket{m|n} = \int_{-\infty}^\infty \Phi^*_m\left(q\right) \Psi_n \left( q - q_e \right) \dev q .
    \label{eq10}
\end{equation}
For distinct ground and excited state potentials, there exists no general analytical expression comparable to equation \eqref{eq2} for equal potentials. Hence, the overlap integral given by \eqref{eq10} needs to be numerically evaluated (see SI note 2 for additional details). \par

\begin{figure}[!htb]
    \centering
    \includegraphics[width=\columnwidth]{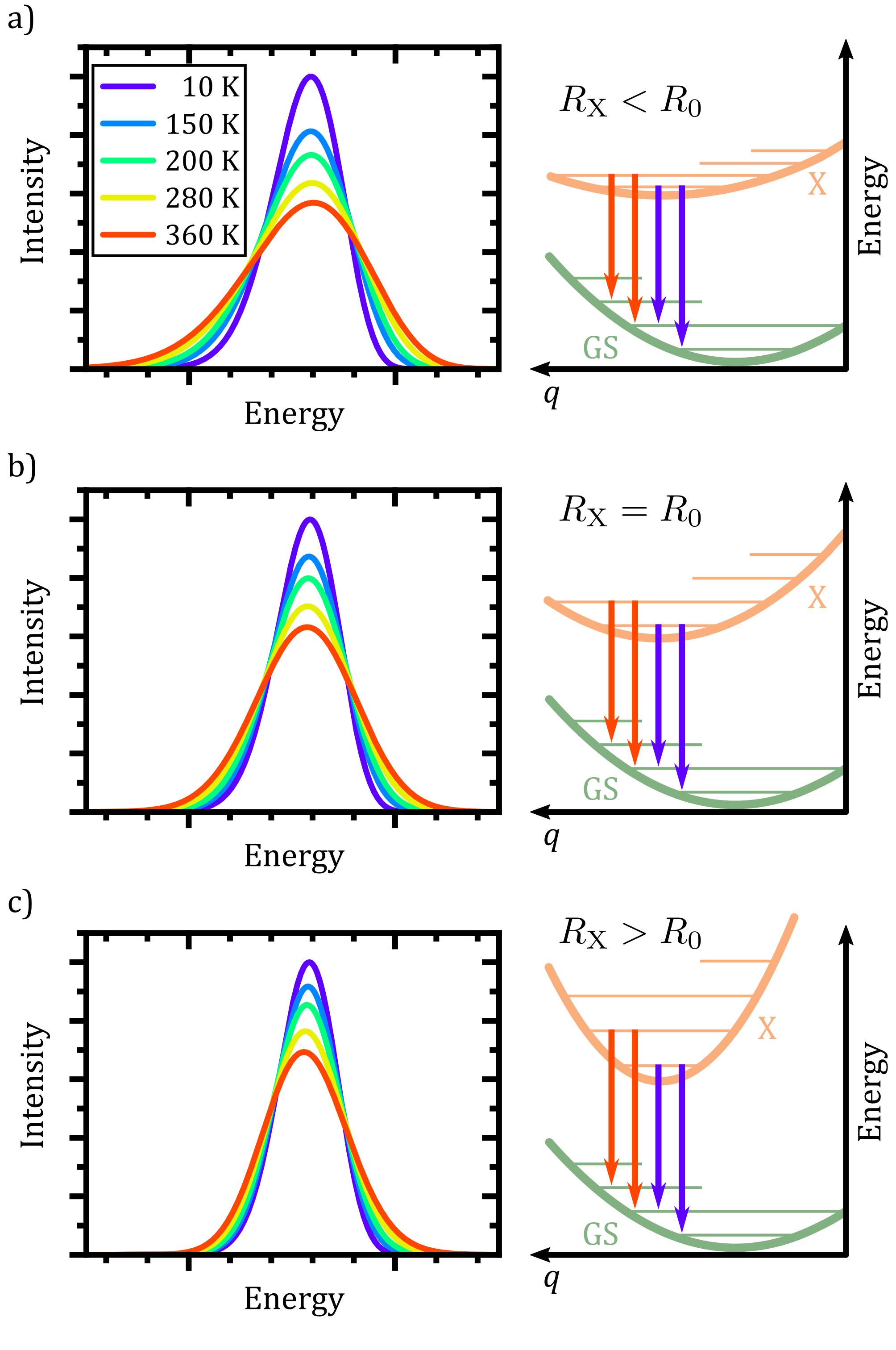}
    \caption{Excimer emission spectra (left row) for an excited state potential shallower (a), equal (b) or steeper (c) than the ground state potential. The schematic PES (right row) show the different harmonic potentials and exemplary vibronic transitions from the X-dimer's vibrational ground (purple) and first excited (red) vibrational states.}
    \label{fig:fig3}
\end{figure}

At this point, it is appropriate to evaluate if and how the emission spectra of non-equal harmonic oscillators deviate from the "ideal" case, i.e. the analytically solvable case of equal ground and excited state potentials. For this purpose we have simulated the emission spectra of three prototypical cases for comparison by means of equations \eqref{eq1} and \eqref{eq10}. The resulting emission spectra $\overline{I}(E,T) \dev E$ are displayed in figure \ref{fig:fig3} for a shallow  ($R_X < R_0$ c.f. Fig. \ref{fig:fig3} a)), an equal ($R_X = R_0$ c.f. Fig. \ref{fig:fig3} b)), and a steep excited state potential ($R_X > R_0$ c.f. Fig. \ref{fig:fig3} c)) at different temperatures.  For the sake of comparability, the ground state potential $R_0$, the X-dimer's energetic and spatial offsets, $D_e$ and $q_e$, as well as the line width parameter $\sigma$ were kept constant while only varying the strength of the excited state potential $R_X = \frac{\mu}{2\hbar^2}E_\text{X,vib}^2$.
At low temperatures, the emission spectra of all three cases show an asymmetric flank towards lower energies. Moreover, an increase in temperature leads to a broadening of the emission spectra while the intensity of the emission maximum decreases as higher vibrational levels of the excited state are populated and hence, intensity from the $\Ket{X,0} \rightarrow \Ket{G,m}$ transitions migrates to vibronic side bands $\Ket{X,n} \rightarrow \Ket{G,m}$. For the case of equal potentials (Fig. \ref{fig:fig3} b)) the asymmetry vanishes with increasing temperature and a Gaussian line shape develops as expected by the semi-classical analytical solution  \cite{Keil_1965}. For the shallow potential of the excited state(Fig. \ref{fig:fig3} a)), the emission profile is highly asymmetric showing a broad tail into the low energy region. For low temperatures, the spectrum's high energy flank is defined by a sharp cut-off which blurs with rising temperature while at the same time, the emission maximum slightly shifts towards higher energies. Moreover, the overall line width is much larger in comparison with the other two cases. The emission spectra in case of the steep excited state potential (Fig. \ref{fig:fig3} c)) have, compared to the other cases, narrow emission profiles, albeit still in the order of \SI{100}{\meV} and show only a slight asymmetry on the high energy flank, becoming more pronounced at higher temperatures. However, the maximum of the emission spectrum shifts to lower energies with rising temperature. Additional details on the peak shifts, the spectral broadening, the asymmetry of the emission spectra as well as computational details are available in the SI as supplementary note 3.
\par 
In conclusion: the spectral features of the X-dimer's temperature dependent emission spectra enable the identification of the relation between the ground and excited state potential and, hence, enable a direct categorization by means of the three cases presented above. 

\section{Analyzing optical spectra}
Following up on the previous theoretical discussion on modeling the luminescence spectra of a multi-molecular excited state, we will now put the developed model in its semi-classical as well as quantum-mechanical form to the test and explore their respective scope of application. For this purpose we chose a CT as well as an excimeric system. We analyze their temperature dependent luminescence spectra, managing both, to evaluate the applicability of our mathematical models as well as extracting relevant parameters of each system. Furthermore, we will discuss the differences between an amorphous system in comparison to a crystalline system for which temperature dependent lattice extensions can no longer be neglected.  

\subsection{CT emission}

\begin{figure*}[!htb]
    \centering
    \includegraphics{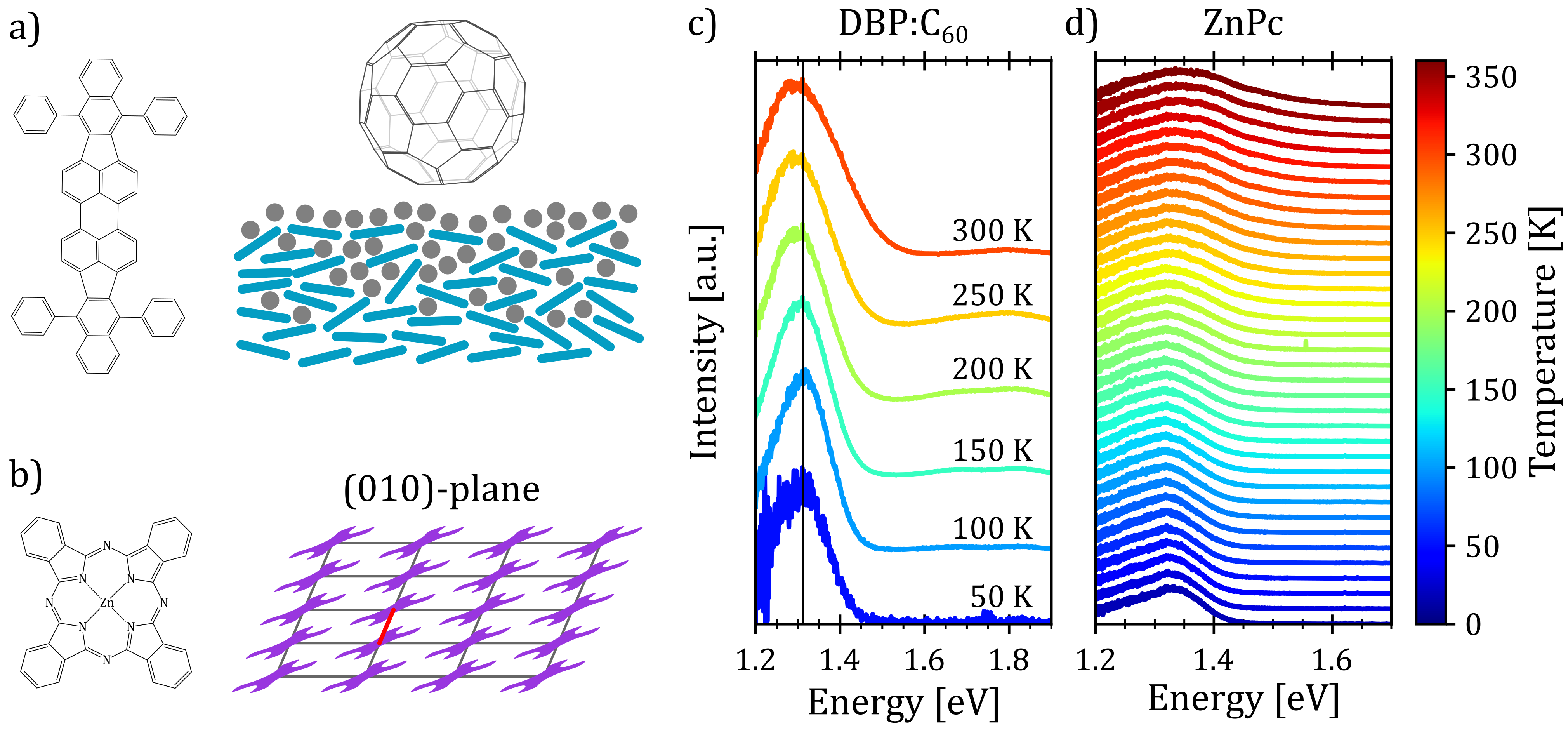}
    \caption{a) Molecular structure of tetraphenyldibenzoperiflanthene (DBP) (left) and fullerene \ce{C60} (top right). A schematic cross section of the DBP-\ce{C60} heterojunction shows the amorphous structure of the DBP (tile lines)-\ce{C60} (grey circles) interface (lower right). b) Molecular structure (left) as well as the (010)-plane molecular packing of the $\alpha$ polymorph (right) of the molecular seminconductor zinc-phthalocyanine (ZnPc). The red line marks the short crystallographic a-direction of the ZnPc $\alpha$ polymorph. c), d) Temperature dependent electroluminecence spectra at an applied voltage of \SI{2}{\volt} of a DBP-\ce{C60} mixed heterojunction (c) and photoluminescence spectra of a crystalline $\alpha$-phase ZnPc thin film (d), respectively. Vertical line in c) serves as guide to the eye. Spectra are offset for clarity.}
    \label{fig:fig4}
\end{figure*}

First, we turn to an amorphous donor-acceptor system comprising an intermixed heterojunction of tetraphenyldibenzoperiflanthene (DBP) as donor and fullerene \ce{C60} as acceptor previously published in  \cite{Linderl_2020}. DBP belongs to the well-known class of perylene dyes. As compared to its parent compound diindenoperylene (DIP), it has an extended $\pi$-conjugated core and, most importantly, four rotatable phenyl rings at the ideno groups at each sideof the perylene core. This changes orientation in thin films from upright standing in the case of DIP to lying down on the substrate for DBP. It was introduced as a novel donor in organic photovoltaics by Fujishima et al. \cite{Fujishima2009} because of its better light absorption due to the favorable molecular orientation. Both, neat films of DBP and its co-evaporated blends with \ce{C60} are amorphous, however, with the mentioned preferential in-plane orientation of the optical transition dipole moment of DBP \cite{Grob2014} (c.f. figure \ref{fig:fig4} a)

Figure \ref{fig:fig4} c) shows the temperature dependent electroluminescence (T-EL) of the DBP-\ce{C60} heterojunction from room temperature down to \SI{50}{\kelvin}. The emission comprises two main features: a pronounced peak at \SI{1.3}{eV} and a weaker one at \SI{1.8}{eV}, the latter growing in intensity with rising temperature. While the emission at higher energies is associated with monomer emission of the donor and acceptor molecules  \cite{Linderl_2020}, the emission feature of interest is the CT emission at \SI{1.3}{\eV}. As evident, the spectra show a broadening as well as a slight red shift of the emission maximum with increasing temperature. Furthermore, we note, that the shape of the spectra also undergoes changes with temperature. For low temperatures, the spectra present an asymmetric shape with a broader flank at the low energy side. With increasing temperature, the emission spectra broaden and the asymmetry slightly shifts towards the high energy side. Hence, the presented phenomenological behavior is consistent with the emission spectra of a X-dimer with a steeper excited state potential compared to the ground state. This qualitative match with the simulated characteristics of an X-dimer encourages us to put the above presented model to a quantitative test. \par
Taking the previous, main findings on the spectral features into account, we can set a parameter space to fit the overall data set within the presented X-dimer model using equation \eqref{eq1}. At first, we assume a Gaussian line shape function $\Gamma$ for each vibronic transition with a temperature independent line width $\sigma$. The emission energy of a single vibronic transition $\ket{X,n} \rightarrow \ket{G,m}$ is given by 
\begin{equation}
    \Delta E_{nm} = D_e + (n+ \frac{1}{2})E_\text{X,vib} - (m+ \frac{1}{2})E_\text{G,vib}.
    \label{eq11}
\end{equation}
From the asymmetry of the emission profile we conclude that the ground and excited state potentials are different, demanding for a numerical calculation of the Franck-Condon factors given by equation \eqref{eq10}. In a first approximation, we assume all other parameters to be independent of temperature, which we justify by the amorphous nature of the sample. Only emission from the excimer's vibrational ground and first four excited states is considered, as the occupation of higher vibrational levels is negligible at these temperatures  \cite{Note_boltzmann_occ}. The fitting routine was carried out by minimizing a single residual function with a least square method by means of a global fit routine provided by the python LMFIT package  \cite{Newville2014}. \par
\begin{figure}[!htb]
    \centering
    \includegraphics{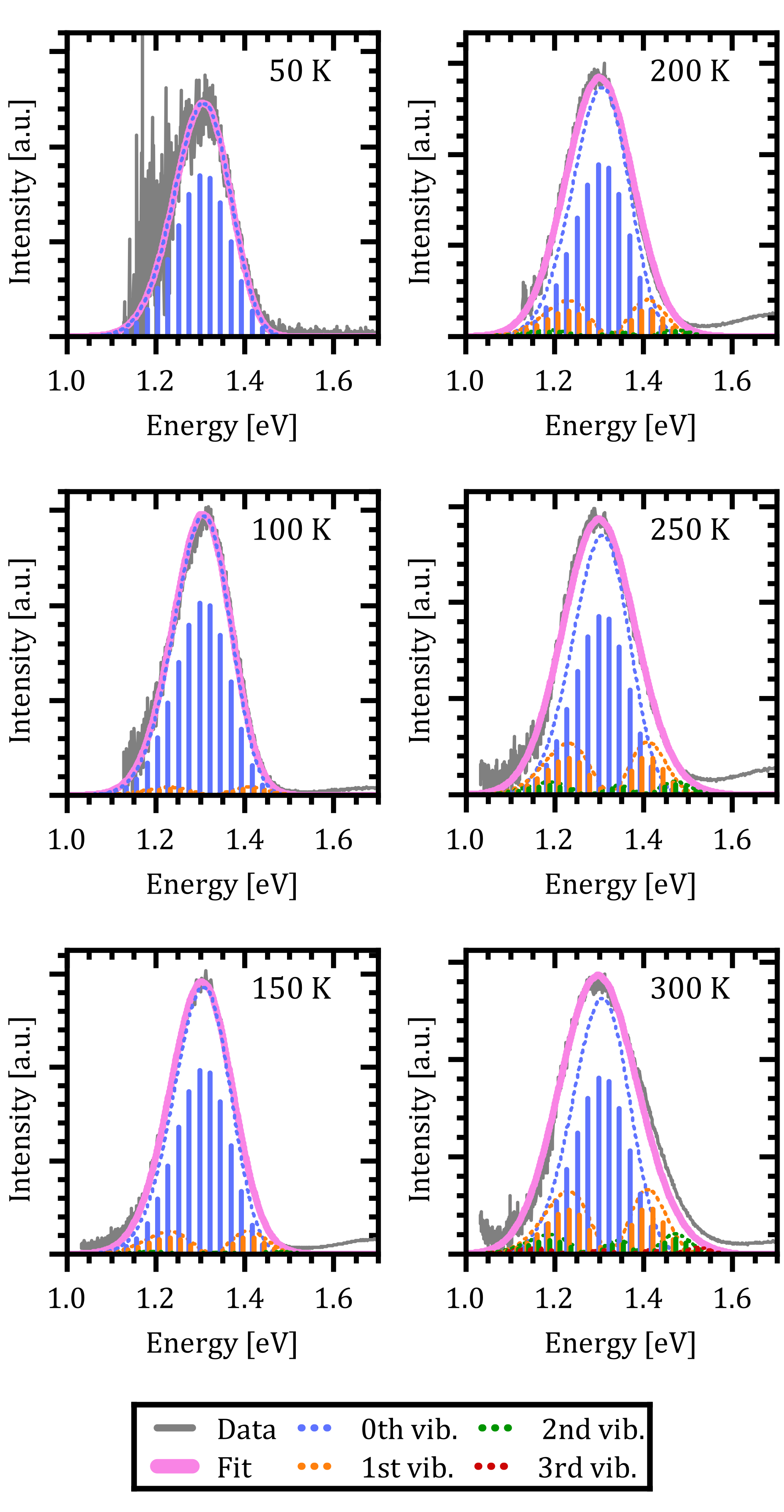}
    \caption{Quantum mechanical simulation of the DBP-\ce{C60} electroluminescence spectra. The legend at the bottom indicates the color code used in the figure for the experimental electroluminescence spectra, the resulting emission spectra (Fit) and the emission from the X-dimer's first four vibrational states.}
    \label{fig:fig5}
\end{figure}
The results are shown in figure \ref{fig:fig5}. The resulting emission spectrum is given by the magenta curve while the experimtenal data is depicted in grey. The envelope emission spectrum is broken down into its individual contributions from the respective X-dimer's vibrational levels which is indicated by the color coding of the individual emission spectra for each level, i.e. blue for vibrational ground state emission, orange for emission from the first excited state, etc. The individual vibronic transitions to the ground state are given by the stick spectra whose heights mark the maximum of the respective Gaussian line shape. The emission spectra are reproduced very well over the full temperature range, despite the use of only a single parameter set. The slight deviation on the high energy side for $T = \SI{300}{\kelvin}$ is attributed to the overlap with the broad monomer emission at around \SI{1.8}{\eV}, which becomes more pronounced as the CT emission broadens with rising temperature.\par

\begin{table}[h]
\small
  \caption{Extracted fit parameters for the CT emission of DBP-\ce{C60}}
  \label{tbl:tab1}
  \begin{tabular*}{0.48\textwidth}{@{\extracolsep{\fill}}llr}
    \hline
    Parameter & & Fitted value\\
    \hline
    X-state vib. energy & $E_\text{X,vib}$ & \SI{27.5(5)}{\milli\eV}  \\
    Ground state vib. energy & $E_\text{G,vib}$ & \SI{23.6(2)}{\milli\eV}   \\
    Energetic offset & $D_e$ & \SI{1.485(2)}{\eV} \\
    Oscillator displacement & $q_e$ & \SI{0.09}{\angstrom}  \\
    Static disorder & $\sigma$ & \SI{13.9(5)}{\milli\eV} \\
    \hline
  \end{tabular*}
\end{table}

The extracted fit parameters are listed in table \ref{tbl:tab1}. First of all, we notice a slightly stronger X-state potential as expected from the qualitative analysis of the evolution of the spectra with temperature. However, for a quantitative assessment of the extracted parameters we need to compare the results to already established analytical models such as Marcus theory  \cite{Vandewal2010, Vandewal2010a} and its derivative including static as well as dynamical disorder by Burke et al.  \cite{Burke2015} performed by Linderl et al. in an earlier publication  \cite{Linderl_2020}. In these approaches the emission spectrum is described by a Gaussian line shape as
\begin{equation}
    \overline{I}(E) \dev E \propto \exp \left(- \frac{[E-(E_\text{CT} -\lambda)]^2}{4\lambda k_\text{b} T} \right)
    \label{eq12}
\end{equation}
where $E_\text{CT}$ is the energetic offset of two equal, displaced harmonic oscillators and $\lambda$ is the effective reorganization energy containing inter- and intra-molecular contributions. In the framework of conventional Marcus theory the line width is solely determined by dynamical disorder making $\lambda$ independent of static disorder. Burke et al. extended the standard approach by substituting the energetic offset $E_\text{CT} = \tilde{E}_\text{CT} + \sigma_\text{B}^2/ 2 k_\text{b} T$ as well as the reorganization energy $\lambda = \tilde{\lambda} - \sigma_\text{B}^2/ 2 k_\text{b} T$ to account for static disorder by introducing the disorder parameter $\sigma_B$. Note that here, the static disorder not only influences the line width of the emission spectra but also the Stokes shift. Combining room temperature electroluminescence and incident photon to current efficiency measurements Linderl et al.  \cite{Linderl_2020} extracted $E_\text{CT,M} = \SI{1.49}{\eV}$ from the Marcus approach and $\tilde{E}_\text{CT,B} = \SI{1.55}{\eV}$, $\sigma_\text{B} = \SI{55}{\milli\eV}$ from the Burke approach for the DBP-\ce{C60} heterojunctions. The corresponding parameter from the X-dimer model is the energetic offset $D_e$ which coincides very well with the all dynamical Marcus model but misses the Burke model by about $\SI{60}{\milli\eV}$. As the X-dimer model accounts for static broadening only on the scale of vibronic transitions and attributes the overall line broadening to an overall dynamical effect, these correspondences are not surprising. Calculating the observed energetic offset from the Burke model by  $E_\text{CT, B} = \tilde{E}_\text{CT,B} - \sigma_\text{B}^2/ 2 k_\text{b} \SI{300}{\kelvin} \approx \SI{1.49}{\eV}$ a good agreement between all three models is achieved. For the second key parameter, the reorganization energy, the Marcus and Burke model yield $\lambda_\text{M} = \SI{0.18}{\eV}$ and $\tilde{\lambda}_\text{B} = \SI{0.12}{\eV}$, respectively. In the X-dimer model, there is no direct analogon, as first of all, by definition the quantum mechanical treatment does not consider a continuous reorganization energy and, second, the excited and ground state have different potentials which excludes a single reorganization energy. However, we can estimate a mean reorganization energy for the hot ground state after emission from the Huang-Rhys parameter  \cite{de_Jong_2015} $S$ via 
\begin{equation}
    \lambda_\text{G} = S E_\text{G,vib} = \frac{1}{2} R_0 q_e^2
    \label{eq13}
\end{equation}
where $R_0$ is the ground state oscillator constant described in equation \eqref{eq3} evaluated using the reduced mass $ \mu = m_\text{DBP} m_{\ce{C60}} /  m_\text{DBP} + m_{\ce{C60}} = \SI{380.243}{\atomicmassunit}$ with \si{\atomicmassunit} being the unified atomic mass unit. This yields $\lambda_\text{G} = \SI{93(3)}{\milli\eV}$, which is lower than the reorganization energy determined from the conventional Marcus approaches. Indeed, an exact match would have been surprising considering the differences between the models. While the above mentioned Marcus based approaches attribute a substantial fraction of the spectral broadening solely to the final state, the X-dimer approach includes the thermal population of X-state low energy vibrational levels, whose emission contribute significantly to the overall line broadening at higher temperatures (c.f. figure \ref{fig:fig5}). From this perspective, Marcus based approaches seem to overestimate the ground state reorganization energy. This becomes even more evident by comparing the extracted reorganization energy with that of the Burke approach. Here, a static energy broadening is assumed, which is the main contribution at low temperatures. This fraction corresponds to the line broadening that is caused by the $\Ket{X,0} \rightarrow \Ket{G,n}$ transitions in the X-dimer model, and hence is mostly determined by the ground state potential and the displacement. Evidently $\sigma_\text{B} = \SI{55}{\meV}$ is much closer to $\lambda_\text{G}$, but, nevertheless, should be interpreted as a dynamic broadening caused by the manifold of energetically tightly spaced vibronic transitions from the vibrational ground state of the excited state to vibrational levels of the electronic ground state. \par
From the above evaluation of our X-dimer model two main conclusions can be drawn: First of all, the temperature dependent spectra can be reproduced over a large temperature range with a single set of parameters, explaining simultaneously the asymmetric line shape at low temperatures, the line broadening with temperature and the red shift of the peak maximum. Second, the comparison to well established models based on Marcus theory confirms similar values for the inherent key parameters of the system, namely the relative energetic position of ground and excited state as well as the line broadening parameter. Even though a one-to-one comparison is not straight forward, the analysis suggests that conventional approaches lead to an overestimation of the reorganization energy of the ground state after population by radiative transitions from the excited state. \par

\subsection{ZnPc excimer emission}
In the next step, we test the described model on a prototypical excimer system, the molecular semiconductor zinc-phthalocyanine (ZnPc) (Fig. \ref{fig:fig4} a ). ZnPc thin films prepared by vacuum sublimation usually adapt the crystallographic $\alpha$-phase with one molecule per unit cell and a P$\overline{1}$ symmetry leading to a slip stack arrangement within the film
plane(c.f. figure \ref{fig:fig4} b)  \cite{erk_2004,Bruder_2010}. The characteristic broad, asymmetric luminescence peak is commonly assigned to an alleged excimer emission \cite{bala09}. We assume the excimer forms along the short crystallographic a-axis (red line in Fig. \ref{fig:fig4} b) ) via convergence of two adjacent monomers, as its characteristic emission is suppressed in the crystallographic $\beta$-phase  \cite{Hammer_2019} as well as in mixtures with fluorinated ZnPc derivatives \cite{Graf_2021} hinting at a sterical hindrance of excimer formation. Recent calculations suggest a strong CT admixture to the lowest excited state in the ZnPc $\alpha$ polymorph \cite{Feng2020, Feng2022} while the lowest excited state is of pure excitonic nature in the crystallographic $\beta$-phase \cite{Feng2022}, corroborating the hypothesis of excimer formation as a preferred relaxation pathway. \par

To compare the semi-classical description with the full quantum mechanical treatment, we first performed temperature dependent photoluminescence (T-PL) measurements on a ZnPc $\alpha$-phase thin film from \SIrange{20}{360}{\kelvin} to obtain a reference data set of excimer emission spectra. The spectra are shown in figure \ref{fig:fig4} d). As can be seen, with increasing temperature the emission shows the typical broadening and the shape of the spectra changes as well. At low temperatures the spectra present an asymmetry towards the low energy tail.
As the temperature rises, the asymmetry shifts towards the high energy flank, indicating a steeper X-dimer potential compared the ground state. Interestingly, the maximum of the spectra shifts towards higher energies with increasing temperature in contradiction with the simulations before. However, so far, we have not included any thermal expansion or contraction of the underlying crystal lattics, as the amorphous DBP-\ce{C60} heterostructure could be fully described without, even though, molecular semiconductors can have up to two orders of magnitude larger expansion coefficients  \cite{ubbelohde_1943, Siegrist_2007, Haas_2007} compared to inorganic compounds \cite{Ashcroft1976, Haynes2014}. A change in the ground state crystal lattice is likely to change the X-dimer's spatial displacement coordinate as well as the ground state potential slope which is common for molecular crystals and can be identified by shifts in phonon frequency with temperature  \cite{Valle2004, Venuti2004, Pinteric_2022}. As these parameters strongly influence the overlap of the nuclear wave functions in the crystalline aggregate, the spectral barycenter as well as the maximum of the emission spectra are highly dependent on $R_0(T)$ and $q_e(T)$ which have to be regarded as functions of temperature and cannot be assumed constant in crystalline materials. \par

Using the same rationale as before, we will first apply the quantum mechanical description to fit the data using equation \eqref{eq1} while estimating the Franck-Condon factors numerically \eqref{eq10} as the asymmetry of the spectra suggests dissimilar ground and X-state potentials. Furthermore, in a first approximation, we assume that the excimer potential is only weakly influenced by the thermal expansion of the crystal lattice. Therefore, the energetic offset to the ground state $D_e$ as well as the X-state potential $R_e$ and, as such $E_\text{X,vib}$, are set to be constant and independent of temperature  \cite{Note_number_free_parameters}. The ground state potential oscillator constant $R_0(T)$ as well as the spatial displacement coordinate $q_e(T)$ are chosen as free parameters. To select a reasonable start value for the X-dimer's vibrational energy quantum we use an independent approach, following the proposed approximation of Birks et al.  \cite{Birks1968}. The full-width-half-maximum (FWHM) at a given temperature $T$ can be approximated by
\begin{equation}
    FWHM(T) = P_0 \coth \left( \frac{E_\text{X,0}}{k_b T} \right)
    \label{eq14}
\end{equation}
from the properties of the Slater sum of a harmonic oscillator \cite{Keil_1965,Birks1968,Amovilli_1991}. Here, $E_\text{X,0}=1/2E_\text{X,vib}$ is the zero-point energy of the inter-molecular excimer vibration and $P_0$ is interpreted as the FWHM at $T=\SI{0}{\kelvin}$ resulting solely from emission from the vibrational ground state. In the semi-classical approximation for equal ground and excited state potentials reported by Keil  \cite{Keil_1965}, or in case the Jacobian $\dev q / \dev E$ does not distort the Gaussian envelope of the oscillator's wave functions $\Psi_n$, as is true for e.g. a linear ground state potential used by Williams and Hebbs \cite{Williams_1951}, equation \eqref{eq14} is the analytical expression for the FWHM of the emission spectra. If the deviation of the line shape of the emission spectra from an overall Gaussian line shape is reasonably small, fitting equation \eqref{eq14} to the temperature dependent FWHM is still acceptable to approximate the excited state potential. We determine $E_\text{X,0}$ to \SI{14.7(4)}{\meV} (supplementary note 4 in SI), which translates to a vibrational energy quantum of $E_\text{X,vib}=\SI{29.4(8)}{\meV}$ that will be used as a starting point for the fit algorithm. As described for the DBP-\ce{C60} CT emission we chose a Gaussian line shape function  with a temperature independent line broadening $\sigma$ as well as the same number of X-state vibrational levels in our simulation. \par

\begin{figure}[!htb]
    \centering
    \includegraphics{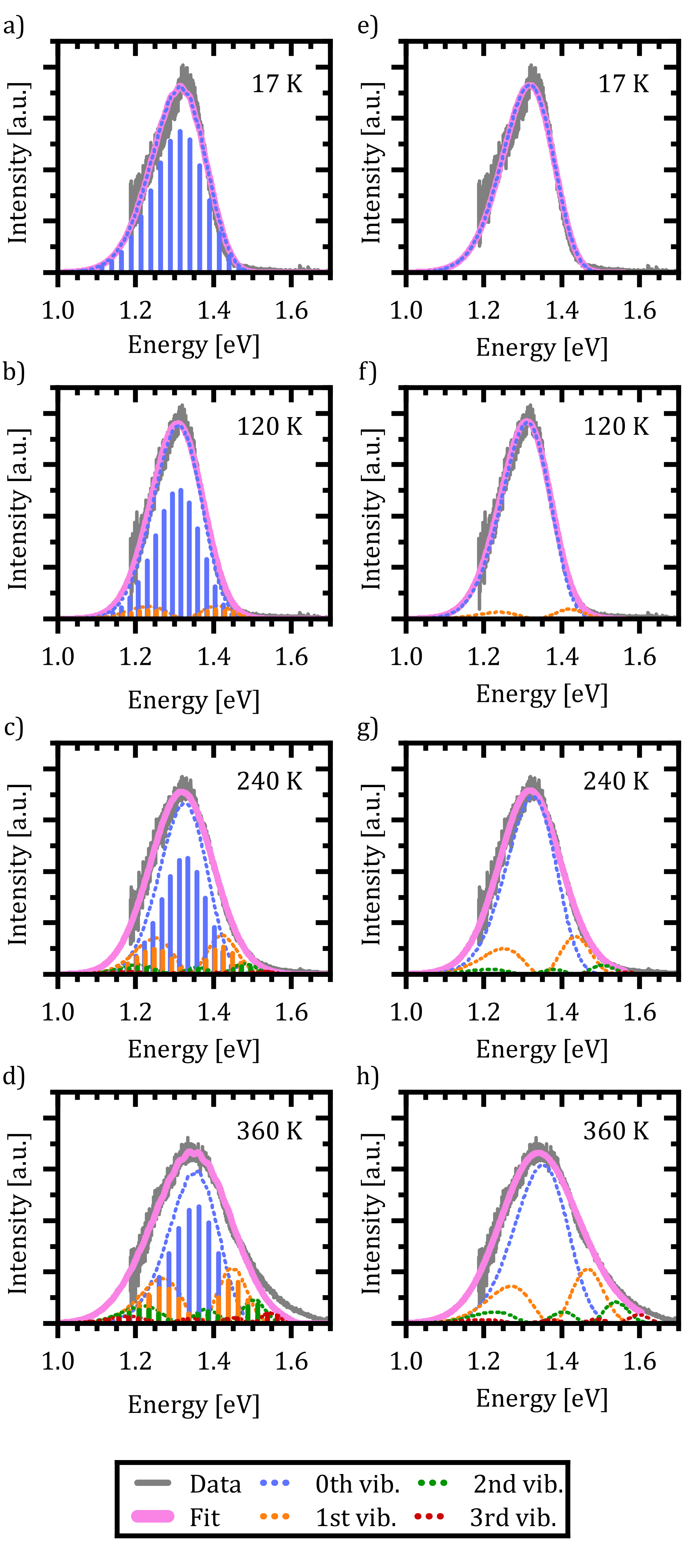}
    \caption{Quantum mechanial (a)-d)) and semi-classical (e)-h)) deduced fits to exemplary ZnPc excimer emission spectra. The legend at the bottom indicates the color code used in each figure for the recorded photoluminescence spectra, the resulting emission spectra (Fit) and the emission from the X-dimer's first four inter-molecular vibrational states.}
    \label{fig:fig6}
\end{figure}

In figures \ref{fig:fig6} a)-d) the resulting spectra and vibrational transitions are shown for exemplary emission spectra at \SIlist{17;120;240;360}{\kelvin} (See SI figure S6 for full data set). The color coding and presentation of the experimental data and the simulated spectra is equal to figure \ref{fig:fig5} where different colors depict emission from different vibrational levels of the X-dimer state and single vibronic transitions are shown as stick spectra. 

Again, the X-dimer model is able to correctly describe the individual experimental emission spectra at a high level. For low temperatures, a slight deviation on the low energy side of the spectra is apparent, leading to a small mismatch between the measured and simulated emission maxima. Yet, the overall line shape is still reproduced very well and with increasing temperature, this mismatch disappears. At \SI{120}{\kelvin} the emission from the first vibrational level becomes apparent, although \SI{94}{\percent} of the intensity are still clustered in the ground state emission. The thermal population of the vibrational levels and, thereby, the increasing contribution to the emission spectra leads to the build up of an asymmetric high energy flank in the emission spectra. At \SI{240}{\kelvin} only \SI{78}{\percent} of the intensity stem from ground state emission while about \SI{17}{\percent} are contributed by the first vibrational level as the second largest contribution. Finally, at \SI{360}{\kelvin} only \SI{58}{\percent} of the emitted photons come from the vibrational ground state, while about a quarter stem from the first and \SI{10}{\percent} from the second vibrationally excited state, while the rest is distributed over higher vibrational levels. At high temperatures $T > \SI{300}{\kelvin}$, the chosen model fails to completely reproduce the high energy flank above \SI{1.5}{\eV}. Notably, the incapacity to fully describe the asymmetric flanks is most pronounced at high temperatures (c.f. fig \ref{fig:fig6} d) \SI{360}{\kelvin}). We attribute this shortcoming to the simplification of deliberately assuming only a single effective inter-molecular vibrational mode. As the thermal population of vibrational modes is the dominant factor determining the spectral envelope  \cite{Warshel_1974}, this simplification may well lead to minor deviations at the extreme ends of the temperature range. Nevertheless, the modelling of the experimental spectra by our dimer approach including a single main mode is still able to reproduce the overall emission spectra while keeping the parameter space for the fitting procedure to a reasonable size. Furthermore, the fitting parameters still enable us to quantify the main parameters determining the potential energy landscape. \par
Using the reduced mass of two ZnPc molecules $\mu = m_\text{ZnPc}/2= \SI{577.916}{\atomicmassunit}/2$ the fit reveals a vibrational energy quantum of $E_\text{X,vib}=\SI{24.7(1)}{\meV}$ for the excimer state. The energetic offset of the excimer potential is estimated to $D_e = \SI{1.49}{\eV}$. Estimating the singlet exciton energy in the crystalline $\alpha$-phase from the lowest Q-band absorption to about \SIrange{1.7}{1.8}{\eV}  \cite{Bruder_2010,Brendel_2015,Graf_2021}, this puts the excimer binding energy in the range of \SIrange{200}{300}{\meV}. As the ground state potential and the X-dimer's spatial displacement are functions of temperature, the absolute values vary (see SI figure S4), and their temperature dependencies are strongly correlated which will be discussed in greater detail below. The vibrational energy quantum of the ground state varies between \SIrange{22}{25}{\meV} putting it quite close to the excited state potential. Here, the highest energies are found for low and high temperatures, while a constant minimum plateau establishes between \SI{100}{\kelvin} to \SI{230}{\kelvin}. The spatial displacement, i.e. the reduction of the distance between two neighboring ZnPc monomers is in the range of \SIrange{0.08}{0.11}{\angstrom} with an contrary behaviour with temperature with respect to the vibrational energy. This is a reasonable result when compared to other excimer systems for which similar values in reduction of inter-molecular distance have been reported  \cite{Birks1968, Warshel_1974, Gierschner2003, Casanova2015, Hoche_2017, Hong2022}. \par
Last but not least, we conduct the above excimer analysis also with the semi-classical approach. For this purpose equation \eqref{eq9} was fitted to each emission spectrum with $R_0$, $q_e$ and $D_e$ as free parameters by a least square method of the SciPy package \cite{virtanen_2020}. Again, to minimize the number of correlated free parameters, we assume the X-state potential as constant with temperature and set it to the fixed value determined from the FWHM using equation \eqref{eq14}. This leads to a slight overestimation of the vibrational energy compared to the results of the quantum-mechanical fit, but, as will be shown below, still yields reasonable results in the reproduction of the emission lineshape and hence gives an opportunity to test the semi-classical approach without any input from the full quantum mechanical model. Six vibrational levels of the excimer were simulated for the emission spectra. \par
Figures \ref{fig:fig6} e)-h) exemplary show the fitted emission spectra at \SIlist{17;120;240;360}{\kelvin} for direct comparison with the quantum mechanical approach (see SI figure S7 for full data set). As before, the resulting lineshape is given in magenta, while the individual contributions of the X-dimer's vibrational levels are color coded. Generic to the semi-classical description, the individual emissions are given by a continuous spectral density rather than discrete individual transitions. The individual vibronic spectra correspond well to the respective envelope spectra from the quantum mechanical calculation shown in figures \ref{fig:fig6} a)-d). Furthermore, despite the higher vibrational energy of the excimer potential, the fractional contributions of the individual vibrational levels to the overall emission intensity are quite similar. At \SI{17}{\kelvin} only the vibrational ground state contributes to the emission. As before, at \SI{120}{\kelvin} the vibrational ground state emission accounts for the largest share of the emission intensity with about \SI{94}{\percent} with the next vibrational level starting to significantly contribute as well. Only at higher temperatures slight deviations to the quantum mechanical model occur. At \SI{240}{\kelvin} the ground state emission amounts to \SI{77}{\percent} and the first vibrational state is at \SI{18}{\percent} while for \SI{360}{\kelvin} still \SI{63}{\percent} of the emission stem from the vibrational ground state followed by the first vibrational level at \SI{24}{\percent}, the second at \SI{9}{\percent} and the rest distributed across higher levers. The fifth excited state only contributes by \SI{0.5}{\percent} at this temperature. Interestingly, the shortcomings in the reproduction of the high temperature spectra are similar to the quantum-mechanical approach, which is plausible if the deviation is caused by the simplification of using just one single vibrational mode. Before discussing the extracted parameters of the excimer's PES, one more limitation of the semi-classical model needs to be addressed, namely the singularity occurring at high emission energies. As already mentioned in the discussion of equation \eqref{eq8}, the energy dependent expression in the denominator of the square-root term leads to a limitation of equation \eqref{eq9} to emission energies below $E_\text{X,n} = D_e + (n + 1/2) E_\text{X,vib}$, i.e. to the singularity of the highest simulated vibrational X-dimer level line shape $E_\text{X,n}-E$. From the quantum mechanical treatment before, we know that the excimeric energy offset $D_e$ is in the order of about \SI{1.5}{\eV}, and hence the simulated emission spectra will be limited to an energy range below $E_\text{X,5} \approx \SI{1.67}{\eV}$. This means that for temperatures above \SI{300}{\kelvin}, the emerging high energy flank of the spectra cannot be simulated completely. This can be recognized as a discontinuity in the fitting parameters $R_0$, $q_e$ and $D_e$ at \SI{330}{\kelvin} where the least square algorithm shifts the emission spectra to higher energies in order to reproduce the experimental data as can be seen in the SI figure S5. \par
Below this temperature, the fitting routine works very well. In fact, the obtained energetic offset as a function of temperature is constant at \SI{1.55(1)}{\eV} without any prior assumption on its behavior with temperature and only \SI{60}{\meV} higher compared to the quantum mechanical model. Moreover, the ground state potential shows a similar behavior with temperature and is characterized by similar parameters between \SI{21.4(3)}{\meV} and \SI{24.5(3)}{\meV} compared to the quantum mechanical model. The same holds true for the spatial displacement with values between \SI{0.10}{\angstrom} and \SI{0.13}{\angstrom}. \par
To summarize, first and foremost, we have proven that we can consistently describe the emission of crystalline $\alpha$-phase ZnPc thin films and its evolution with temperature with the presented X-dimer model based on a single inter-molecular vibrational mode. Hence, our results strongly suggest that the characteristics of the photo- and electroluminescence \cite{Hammer_2019} are indeed the result of an excimer formation located \SI{200}{\meV} to \SI{300}{\meV} below the free exciton. The lack of free exciton luminescence suggests that there is no energetic barrier between the two states, leading to an exothermic instantaneous population of the excimer state. As the dominant mechanism we suggest a convergence of two adjacent molecules, reducing the inter-molecular distance by about \SI{0.1}{\angstrom}. Our results further suggest, that especially at high temperatures, a second vibrational mode is likely to contribute to the emission as well. By means of the overall comparison, we showed that a semi-classical approach can be well suited to describe the measured excimer emission as long as the inherent limitations are carefully considered.

\FloatBarrier
\section{Discussion}
Regardless of its successful application to two inherently different molecular systems, in the following, we will discuss potential pitfalls and shortcomings of the proposed X-dimer model, as well as strategies to overcome those. Applying the X-dimer model  to extract the material inherent parameters without any additional information on the ground and X-state potentials can be challenging and obtained results should be treated cautiously. The parameters defining the ground and the excited state potential as well as the oscillator displacement are highly correlated as they mutually contribute to the spectral broadening, which is indicated by their high co-variance values. In conventional Franck-Condon based models this drawback is usually circumvented by combining these three parameters in the Huang-Rhys parameter which then effectively describes the spectral broadening by a single parameter. However, this approach fails for materials where the asymmetry in line shape, especially at higher temperatures, refers to a different ground state and excited state potential. Nevertheless, there exist strategies to reduce the free parameter space and to control these uncertainties for certain cases. For example in the above presented analysis of the ZnPc excimer, the experimental data allows for an estimation of the X-dimer's zero point energy by means of the semi-classical approximation of the FWHM \eqref{eq14} which can be either used as a reliable start value for the fitting procedure, as demonstrated for the quantum mechanical approach, or even eliminate the parameter at all by assuming it to be a fixed value, as shown for the semi-classical approach. Furthermore, as demonstrated by Birks et al.  \cite{Birks1968} the low temperature spectra allow to infer the ground state potential parameter $R_0$ and the X-state displacement $q_e$, because the low temperature emission spectra are dominated by transitions from the X-dimer's vibrational ground state to ground state phonons. As a consequence, a thorough evaluation of the data set taking the limits of the underlying model into account has to be performed before detailed analysis. \par
With these conditions in mind, it is however possible to gain considerable insights into the photophysical processes of multi-molecular excited states by directly evaluating the temperature dependent spectrally resolved steady state luminescence. As demonstrated on different material systems the one dimensional PES can be extracted, that reveals, within the discussed limitations, the reorganization energies after photo emission, the inter-molecular displacement as well as the binding energies. 

\section{Conclusion}
In this contribution we presented the X-dimer model to analyse the temperature dependence of luminescence spectra of multi-molecular excited states originating from Franck-Condon like transitions between inter-molecular vibrational levels. We evaluated this model for two different material systems, the CT emission of the amorphous donor-acceptor bulk heterojunction DBP-\ce{C60} and the excimer emission of $\alpha$-ZnPc crystalline thin films. The emission of both systems was reproduced very well by the presented model and the extracted parameters agree with that of established benchmark models. \par
From our analysis we conclude that there is no need to include a significant contribution by static disorder to explain broad spectral line shapes even at low temperatures, which seems reasonable for amorphous systems but appears rather counter intuitive for crystalline aggregates such as $\alpha$-ZnPc. Rather, our results suggest that a manifold of vibronic transitions between low energy inter-molecular vibrations is indeed sufficient to describe the emission over a large temperature range and, hence, supports the picture of \textit{dynamic broadening} as suggested previously in various forms \cite{Tvingstedt_2020, Vandewal2010, Vandewal2010a, Benduhn2017, Goehler2021}.\par
In summary, by the two seminal material examples and the information gained by applying our X-dimer model, we want to encourage a more extended analysis of emission spectra which by their lineshape and its temperature dependence indicate a multi-molecular origin.

\section*{Methods}
\paragraph*{Sample preparation \& characterization}
\textit{DBP-\ce{C60}} devices have been prepared via thermal evaporation on a \SI{30}{\nm} prepared poly(3,4-ethylenedioxythiophene):poly(styrenesulfonate)(PEDOT: PSS) coated indium-tin-oxide (ITO) covered glass substrate. The \SI{50}{\nm} active layer (DBP:\ce{C60}, molar ratio 1:2) is enclosed by neat \SI{5}{\nm} DBP donor and \SI{10}{\nm} \ce{C60} layers. A \SI{5}{\nm} bathocuprione (BCP) exciton blocking layer is added on top to prevent quenching at the \SI{100}{\nm} aluminium (Al) cathode. The total layer structure is ITO/PEDOT:PSS/DBP/DBP:\ce{C60}/\ce{C60}/BCP/Al. Morphological characterization confirming the amorphous nature of the donor acceptor bulk heterojunction as well as a detailed description of the sample preparation is given in  \cite{Linderl_2020}. \par
\textit{ZnPc thin films} were prepared via vacuum sublimation of ZnPc, purified by twofold gradient sublimation, at a base pressure of \SI{e-9}{\milli\bar} while controlling the deposition rate and film thickness by a quartz crystal microbalanace. A film of \SI{30}{\nm} was evaporated on a \SI{200}{\nm} thick thermal oxide layer on top of a silicon wafer at a rate of \SI{10}{\angstrom\per\minute}. These crystalline films grow in the thermodynamically metastable $\alpha$-phase and are composed of small crystallites (lateral dimensions about \SIrange{20}{50}{\nm}) with a high crystallinity along the out-of-plane direction (as large as the film thickness as determined by X-ray diffraction analyses)  \cite{Hammer_2019}. 

\paragraph*{Optical measurements}
\textit{Temperature dependent electroluminescence measurements} are described in  \cite{Linderl_2020}. Specifically, the spectra were recorded using a liquid-nitrogen-cooled CCD camera (PyLoN:100BR\_ eXcelon, Princton Instruments) coupled with a spectrometer (SP2300i, Princton Instruments) with spectral sensitivity in the wavelength range of approximately 300-1000\,nm. The measurements were performed under constant dc voltage drive (2\,V in all cases shown in this work) from a Keithley source meter. The samples were transfered into a liquid-helium-cooled cryostat (Cryovac) with an inert gas atmosphere (approximately 300\,mbar He) without air exposure.
\par
\textit{Temperature dependent photoluminescence measurements} on ZnPc thin films were performed at \SI{685}{\nm} cw-excitation with a Coherent OBIS LS/LX solid state laser. The sample was mounted on a cold finger in a CryoVac helium cryostat with a silicon diode as temperature sensor located at the sample position for reliable temperature measurements and control. The laser was focused onto the sample via a OLYMPUS SLMPLNx 50 ($NA = \num{0.35}$, 50x magnification) long working distance objective. The luminescence signal was captured through the same objective and guided to a Princeton Instruments Acton SP-2500i spectrometer with a PIXIS 100BR eXcelon peltier cooled CCD camera. Spectra were recorded at each temperature with an acquisition time of \SI{60}{\s} after a \SI{15}{\minute} delay to ensure thermal equilibrium. The spectra have been background corrected before analysis.

\section*{Code availability}
We provide the xDimer python package containing the functions to simulate luminescence spectra within the semi-classical and quantum-mechanical X-dimer model on a Zenodo repository at \href{https://doi.org/10.5281/zenodo.6707037}{https://doi.org/10.5281/zenodo.6707037}.

\section*{Acknowledgements}
S.H. gratefully acknowledges funding from the German Research Foundation (DFG) through the project 490894053.
Financial support by the Bavarian State Ministry of Science, Research and the Arts within the collaborative research network "Solar Technologies go Hybrid (SolTech)" is gratefully acknowledged by WB and JP. JP acknowledges funding from the German Research Foundation (DFG) within the project PF385/12-1. WB acknowledges funding from the German Research Foundation (DFG) within the project BR 1728/14-2.

\FloatBarrier
\balance
\printbibliography[title=Notes and references] 
\end{document}


\title{Supplementary Information: \\ 
Spectroscopic analysis of vibrational coupling in multi-molecular excited states}
\author[1,4]{Sebastian Hammer\thanks{mail to: sebastian.hammer@mail.mcgill.ca}}
\author[2]{Theresa Linderl}
\author[1]{Kristofer Tvingstedt}
\author[2]{Wolfgang Brütting}
\author[1,3]{Jens Pflaum\thanks{mail to: jens.pflaum@physik.uni-wuerzburg.de}}

\affil[1]{Experimental Physics VI, Julius Maximilian University Würzburg, 97074 Würzburg, Germany.}
\affil[2]{Institute of Physics, University of Augsburg, 86135 Augsburg, Germany.}
\affil[3]{Bavarian Center for Applied Energy Research, 97074 Würzburg, Germany.}
\affil[4]{Present address: Center for the Physics of Materials, Department of Physics and Department of Chemistry, McGill University, Montreal, QC H3A 2K6, Canada.}

\date{} 

\maketitle

\newpage

{\centering\textbf{Supplementary note 1: Semi-classical model of excimer emission}}

As described in the main manuscript, the probability of the X-dimer geometry to adopt the displacement $q^*$ is given by the probability density $\abs{\Psi(q^*)}^2\dev q$ which is directly related to the emission spectrum $\overline{I}(E)\dev E$ for a photon energy $E$ by 
\begin{equation}
    \overline{I} \left( E \left( q^* \right) \right) \dev E = \abs{ \Psi \left(q^* \right) }^2\dev q.
    \label{S1}
\end{equation}
Furthermore, the spatial coordinate $q$ can be expressed as a function of the photon energy $q(E)$, and with an appropriate Jacobian for the coordinate transformation the emission spectrum for a transition from an excited state $\Ket{X,n}$ to the electronic ground state reads
\begin{equation}
   \overline{I}_n\left( E \right) \dev E = \abs{ \Psi_n \left(q \left( E \right) \right) }^2 \frac{\dev q}{\dev E} \dev E
    \label{S2}
\end{equation}
with the wave function $\Psi_n$ as the wave function of the vibrational state $\Ket{n}$. This expands the semi-classical temperature dependent emission spectrum as given by equation (9) in the manuscript by
\begin{equation}
    \begin{split}
        \overline{I} \left(E, T \right) \dev E &= \sum_n  P \left(n,T \right) \overline{I}_n\left( E \right) \dev E  \\
        &= \sum_n  P \left(n,T \right) \abs{ \Psi_n \left(q \left( E \right) \right) }^2 \frac{\dev q}{\dev E} \dev E. 
        \label{S3}
    \end{split}
\end{equation}\par

To establish an analytical relation between the spatial coordinate and the photon energy, we use equations (3), (5) and (6) from the main text to calculate the photon energy and define the respective spatial coordinate for a vibrational X-dimer state $\Ket{n}$ defined by equation (4) in the main text to
\begin{equation}
    \Tilde{q}_n \left( E\right) = \sqrt{\frac{E_\text{X,n}-E}{R_0}}-q_e 
    \label{S4}
\end{equation}
yielding 
\begin{equation}
\frac{\dev \Tilde{q}_n}{\dev E} = \frac{1}{2\sqrt{R_0 \left(E_\text{X,n}-E\right)}}
    \label{S5}
\end{equation}
as the Jacobian. Using the wave functions of the quantum mechanical harmonic oscillator, see e.g. \cite{Parson2007},
\begin{equation}
\psi_n \left( \Tilde{q}_n \right) = \left( \frac{\alpha}{\pi} \right)^\frac{1}{4} \frac{1}{\sqrt{2^n n!}} H_n \left( \sqrt{\alpha} \Tilde{q}_n \right) \exp\left(-\frac{1}{2}\alpha \Tilde{q}_n^2\right)
    \label{S6}
\end{equation}
with the respective hermite polynome $H_n (x)$ and an oscillator parameter $\alpha = \mu E_\text{X,vib}/ \hbar^2$ the emission spetrum $\overline{I}_n\left( E \right) \dev E$ defined in equation \eqref{S2} can be expressed analytically using equations \eqref{S4}-\eqref{S6}. \par
Evaluating \eqref{S2} for the emission $\Ket{X,n} \rightarrow \Ket{G}$ from the first six vibrational levels $n \in \lbrace 0,1,2,3,4,5 \rbrace$ yields

\begin{align}
\overline{I}_0(E) \dev E &= \frac{1}{2} \sqrt{\frac{\alpha}{R_0 \pi \left(E\mi{X,0} -E\right)}}\, \exp \left( -\alpha \tilde q_0^2\right) \dev E \label{S7}\\
\overline{I}_1(E) \dev E &= \sqrt{\frac{\alpha}{R_0 \pi \left(E\mi{X,1} -E\right)}} \, \alpha \tilde q_1^2 \, \exp \left( -\alpha \tilde q_1^2\right) \dev E \label{S8}\\
\overline{I}_2(E) \dev E &= \frac{1}{4} \sqrt{\frac{\alpha}{R_0 \pi \left(E\mi{X,2} -E\right)}}\, \left( 2\alpha \tilde q_2^2-1 \right)^2 \, \exp \left( -\alpha \tilde q_2^2\right)\dev E \label{S9}\\
\overline{I}_3(E) \dev E &= \frac{1}{6} \sqrt{\frac{\alpha}{R_0 \pi \left(E\mi{X,3} -E\right)}} \, \left(2\alpha \tilde q_3^2 -3 \right)^2  \alpha \tilde q_3^2  \, \exp \left( -\alpha \tilde q_3^2\right)\dev E \label{S10}\\
\overline{I}_4(E) \dev E &= \frac{1}{48} \sqrt{\frac{\alpha}{R_0 \pi \left(E\mi{X,4} -E\right)}} \, \left( 4 \alpha ^2 \tilde q_4^4 - 12 \alpha \tilde q_4^2 +3\right)^2 \, \exp \left( -\alpha \tilde q_4^2\right)\dev E \label{S11}\\
\overline{I}_5(E) \dev E &= \frac{1}{120} \sqrt{\frac{\alpha}{R_0 \pi \left(E\mi{X,5} -E\right)}} \, \left( \left( 2 \alpha \tilde q_5^2 -5 \right)^2 -10\right)^2 \alpha \tilde q_5 ^2 \, \exp \left( -\alpha \tilde q_5^2\right)\dev E 
\label{S12}
\end{align}

\newpage
{\centering\textbf{Supplementary note 2: Numerical evalution of the quantum-mechanical overlap intergral}}

The $\propto \exp \left( -q^2 \right)$ dependency of the oscillator wave functions \eqref{S6} enables a straightforward numerical integration applying fixed suitable boundaries as the wave function only slightly extends beyond the turning points of the classical oscillator at any given energy. Hence, equation (10) of the main manuscript simplifies to 
\begin{equation}
        \Braket{m|n} = \int_{-\Delta q}^{\Delta q} \Phi^*_m\left(q\right) \Psi_n \left( q - q_e \right) \dev q .
    \label{S13}
\end{equation}
The integration interval $[-\Delta q, \Delta q]$ has to be chosen as such, that it almost fully includes the ground state wave function $\Phi_m\left(q\right)$ located around $q=0$ and the excited state wave function $\Psi_n \left( q - q_e \right)$ at $q=q_e$. Suitable integration boundaries can be estimated using the classical turning points of a harmonic oscillator with oscillator constant $R$ as
\begin{equation}
    q = \sqrt{\frac{2\left(n+\frac{1}{2}\right)E_\text{vib}}{R}}
    \label{S14}
\end{equation}
where $E_\text{vib}$ represents the vibrational energy quantum. For an exemplary reduced mass of $\mu = \SI{350}{\atomicmassunit}$ and a vibrational energy quantum $E_\text{vib}= \SI{20}{\meV}$ equation \eqref{S13} yields  $q (n=5) = \SI{0.18}{\angstrom}$, $q (n=10) = \SI{0.22}{\angstrom}$ and $q (n=25) = \SI{0.34}{\angstrom}$. For a common spatial displacement of $q_e \approx \SI{0.2}{\angstrom}$ this means that for integration boundaries of $\Delta q = \pm \SI{0.5}{\angstrom}$ even excited state wave functions as high as $n=10$ are almost completely included while the ground state wave functions are included even up to $n=25$. The numerical integration is performed using the "simpson" integration function within the SciPy python package  \cite{virtanen_2020}.

\par 
Of course, equation (10) in the main manuscript has to be evaluated for all transitions $\Ket{X,n} \rightarrow \Ket{G,m}$ contributing to the emission spectrum at temperature $T$ and hence the maximum evaluated vibrational levels of the ground and X-Dimer state, $M:=\lbrace 0,1,\dots, m_\text{max} \rbrace$ and $N= \lbrace 0, 1, \dots, n_\text{max} \rbrace$, respectively, have to be chosen accordingly. This results in a total of $\abs{N \times M}$ individual transitions which need to be calculated. To minimize evaluation time during fit procedures, $n_\text{max}$ should be chosen according to the excepted vibrational energy quantum of the X-dimer state and the highest temperature. For example, as indicated in note 53 in the main manuscript, for a vibrational energy quantum of \SI{25}{\meV} and a temperature of \SI{400}{\kelvin} the population probability of the 5th vibrationally excited state is \SI{1.4}{\percent}.

\newpage
{\centering\textbf{Supplementary note 3: Computational details for simulated emission spectra in figure 3}}

The emission spectra presented in figure 3 in the main manuscript have been generated by assuming a reduced mass of $\mu = \SI{288.958}{\atomicmassunit}$ (corresponding to the reduced mass of a zinc-phthalocyanine dimer). The constant parameters where chosen as $E_\text{vib,G} = \SI{25}{\meV}$ for the ground state vibrational energy, $D_e = \SI{2}{\eV}$ as the energetic offset, $q_e = \SI{0.1}{\angstrom}$ as the spatial offset and $\sigma = \SI{20}{\meV}$ as the line width parameter of the gaussian line shape function. For the steep and shallow X-dimer state we set $E_\text{vib,X} = \SI{30}{\meV}$ and $E_\text{vib,X} = \SI{20}{\meV}$, respectively. Emission spectra where simulated in \SI{5}{\kelvin} steps from \SIrange{5}{400}{\kelvin} and in \SI{25}{\kelvin} steps between \SI{400}{\kelvin} and \SI{1000}{\kelvin} considering 100 vibrational levels for the ground state oscillator and 20 vibrational levels for the X-dimer state. The Boltzmann population has been numercially evaluated assuming $z=100$ vibrational states of the X-dimer. The numerical integration has been performed within $\Delta Q = \pm \SI{3}{\angstrom}$. \par
\begin{figure}[h]
    \centering
    \includegraphics{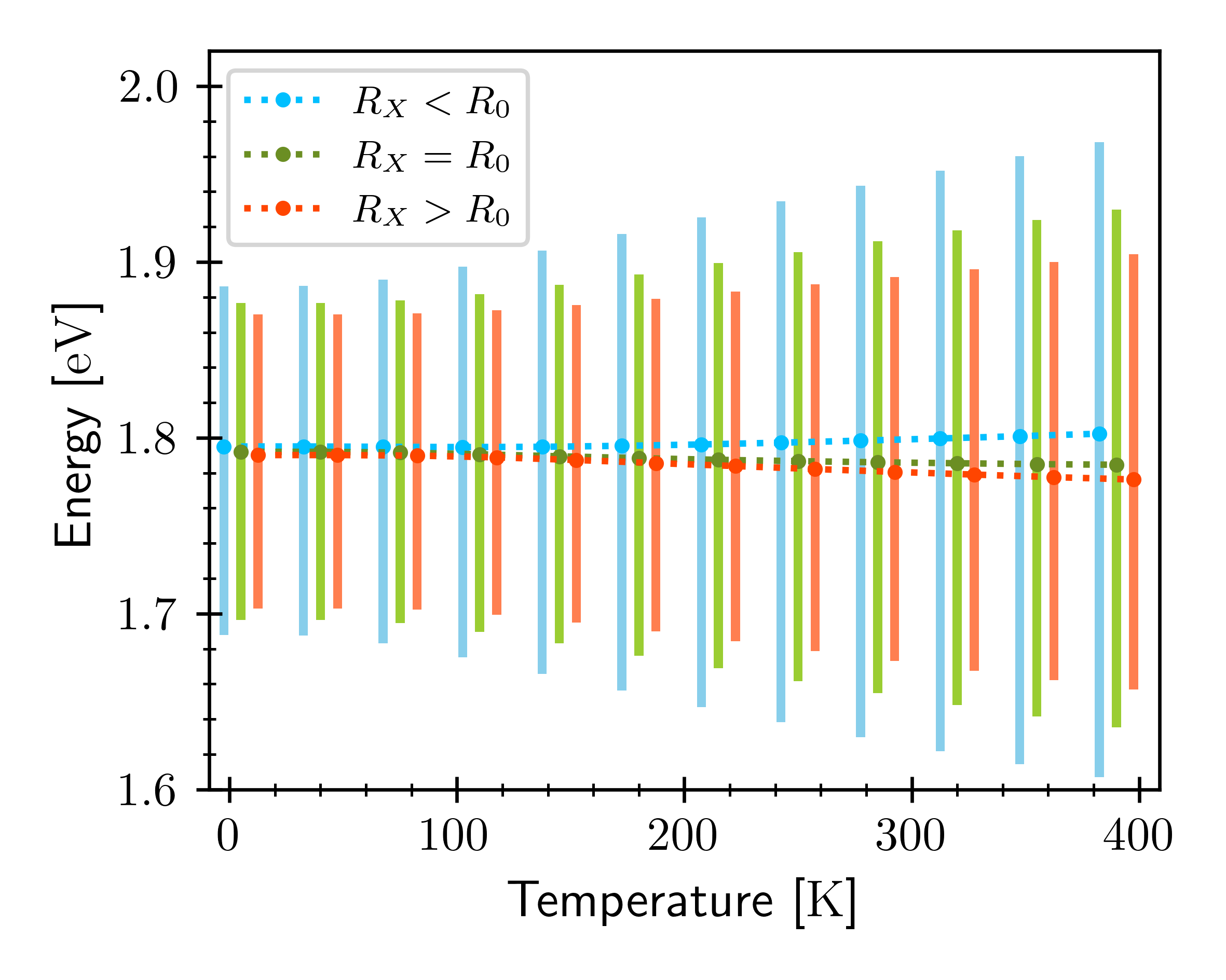}
    \caption{Maximum peak position (dots) and the HWHM to the low and high energy side (bars) with temperature for all three cases. Dotted lines as guide-to-the-eye for the peak shift.}
    \label{fig:S1}
\end{figure}

Figure \ref{fig:S1} shows the peak maximum (dots) as well as the asymmetric half-width-half-maximum (HWHM) towards the low and high energy side of the spectrum (bars) for selected temperatures between \SI{5}{\kelvin} and \SI{400}{\kelvin} illustrating the broadening of the spectra as well as the maximum peak shift with temperature.

\begin{figure}[h]
    \centering
    \includegraphics{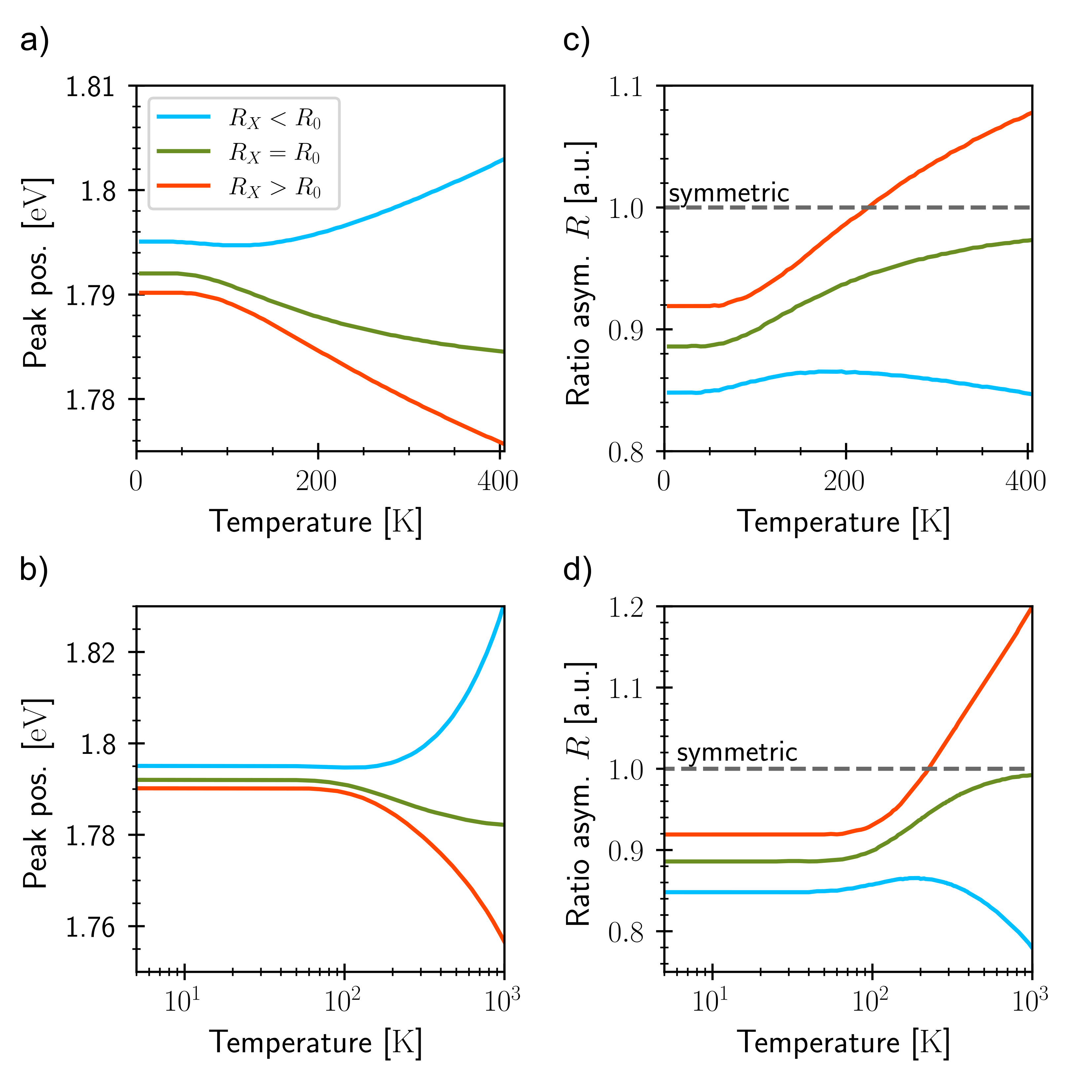}
    \caption{Maximum peak positions (a, b) as well as ratio of the low and high energy side HWHM (c,d) as a function of temperature. A ratio of one indicates a symmetric emission profile, while values smaller/larger than one indicate an asymmetry towards the low/high energy side.}
    \label{fig:S2}
\end{figure}

\FloatBarrier
The evolution of the peak shift with temperature is also shown in figures \ref{fig:S2} a) and b) for different temperature ranges indicating the different trends towards higher and lower emission energies for all three cases of potential strengths. To provide deeper insight into the temperature dependence of the asymmetry between high and low energy flank of the emission spectra figures \ref{fig:S2} c) and d) depict the ratio between the HWHM of low and high energy flank of the simulated emission spectra. The ratio is calculated as
\begin{equation}
    R = \frac{HWHM_{\text{low energy}}}{HWHM_{\text{high energy}}}.
    \label{S15}
\end{equation}
Hence, a value of $R=1$ indicates a symmetric emission profile while values of $R<1$ and $R>1$ indicate an asymmetry towards the low and high energy flank, respectively. For the case of the weaker excited state potential we see $R<1$ over the whole temperature range showing the asymmetry towards the low energy side, while for the stronger excited state potential the assymetry changes from the low to the high energy flank with rising temperature. For the case of the equal ground and excited state potential $R$ converges towards 1 with rising temperature, which means the spectrum strives towards a symmetric emission profile. 

\newpage
{\centering\textbf{Supplementary note 4: Estimation of the zero point energy from the FWHM temperature  of the $\alpha$-ZnPc emission spectra }}

The FWHM of the temperature dependent $\alpha$-ZnPc emission spectra has been extracted by the abscissa difference between the half-intensity point of the low and high energy flank of a smoothed spectral curve. The errors are estimated from the noise of the spectral curve. The data was fitted using equation (14) from the main manuscript yielding the fit parameters $S_0 = \SI{147.9(14)}{\meV}$ and $T_0 = \frac{E_\text{X,0}}{k_b} = \SI{171.4(52)}{\kelvin}$.

\begin{figure}[h]
    \centering
    \includegraphics{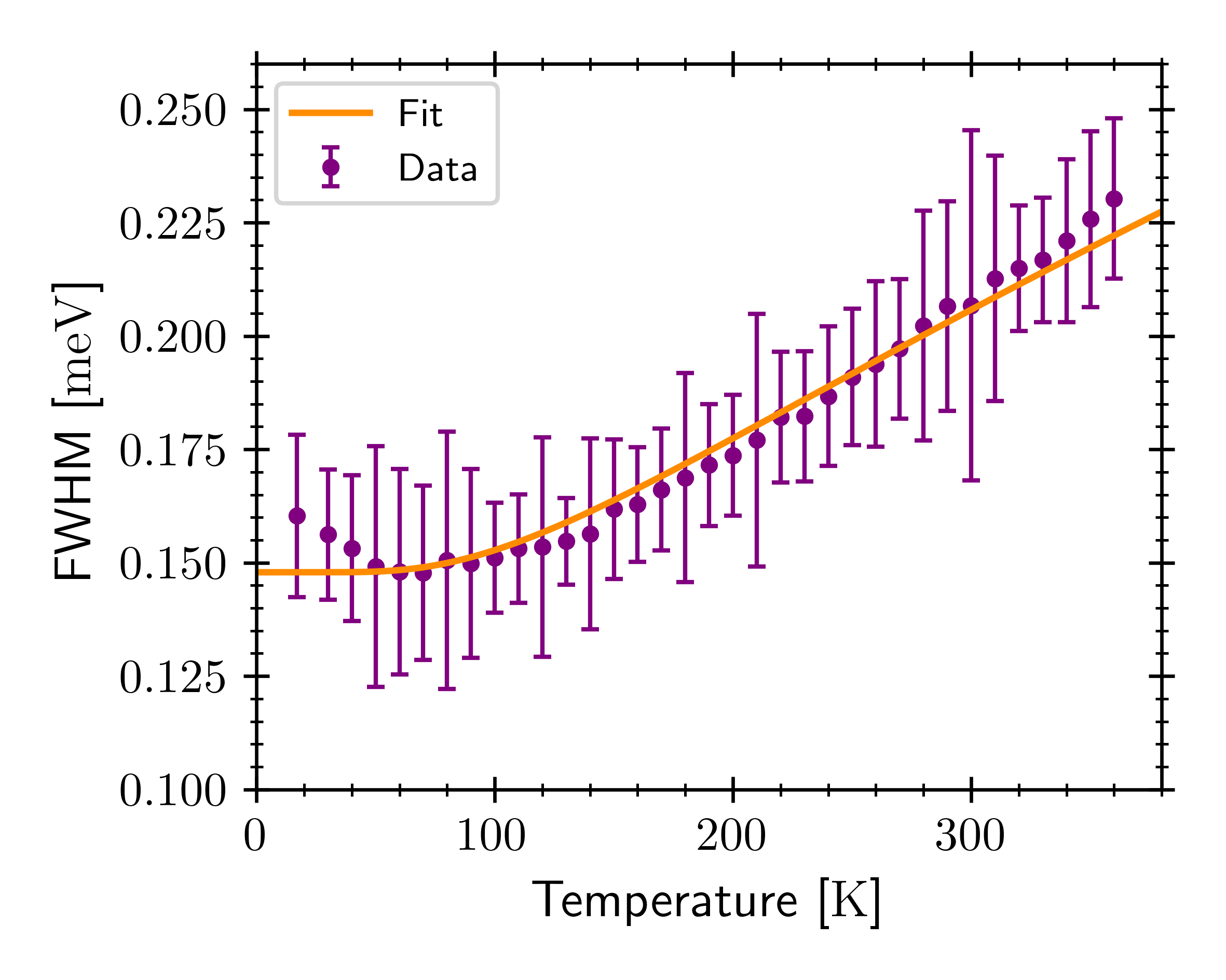}
    \caption{ZnPc FWHM with temperature together with fit curve from equation (14).}
    \label{fig:S3}
\end{figure}

\newpage
{\centering\textbf{Supplementary note 5: ZnPc emission: X-dimer model}}

\begin{figure}[h]
    \centering
    \includegraphics{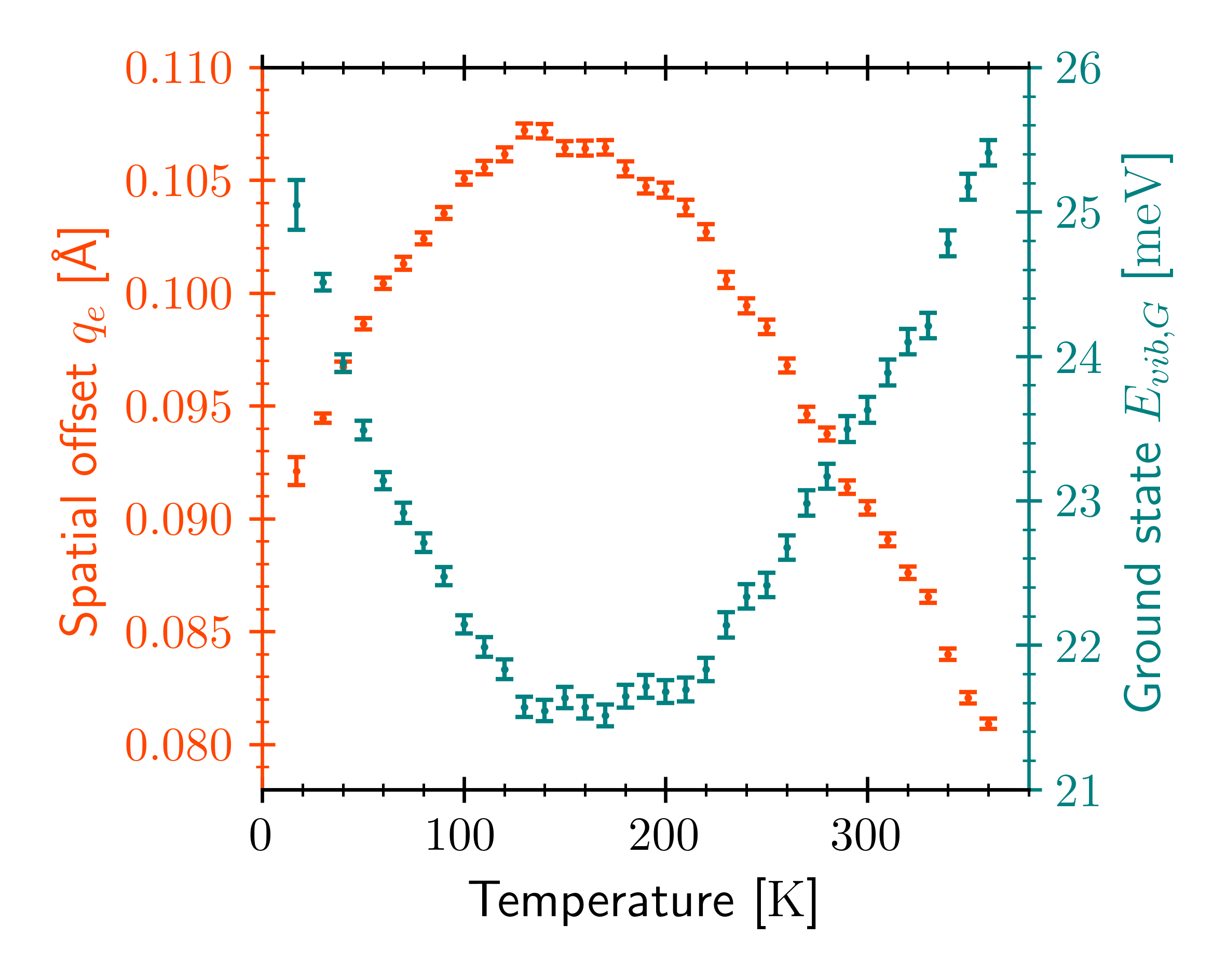}
    \caption{Temperature dependence of spatial offset $q_e$ (orange, left y-axis) and vibratioanl energy of the ground state $E_\text{vib,G}$ (teal, left y-axis) extracted from the quantum mechanical X-dimer fit to the ZnPc luminescence data.}
    \label{fig:S4}
\end{figure}

\begin{figure}[h]
    \centering
    \includegraphics{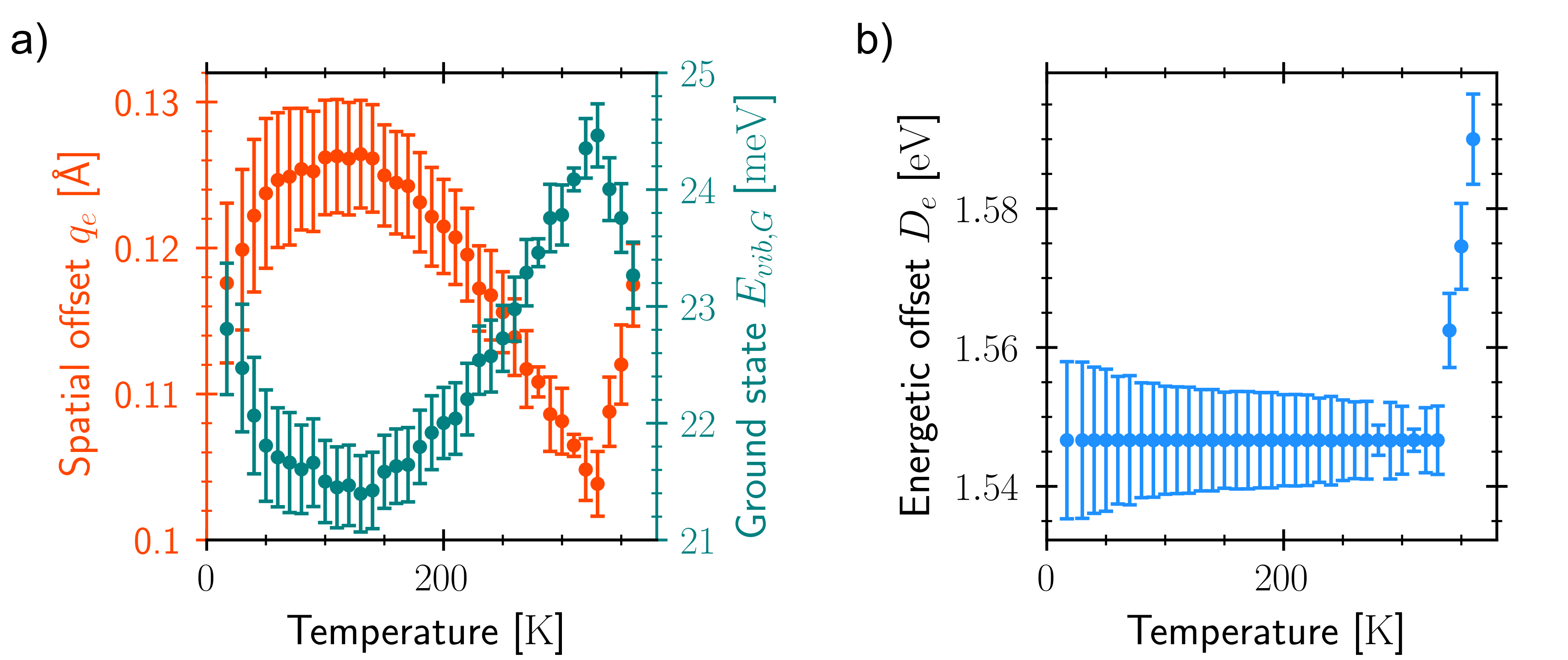}
    \caption{Temperature dependence of spatial offset $q_e$ (orange, left y-axis) and vibrational energy of the ground state $E_\text{vib,G}$ (teal, left y-axis) (a) as well as energetic offset $D_e$ (b) extracted from the semi-classical X-dimer fit to the ZnPc luminescence data.}
    \label{fig:S5}
\end{figure}

\begin{figure}[h]
    \centering
    \includegraphics{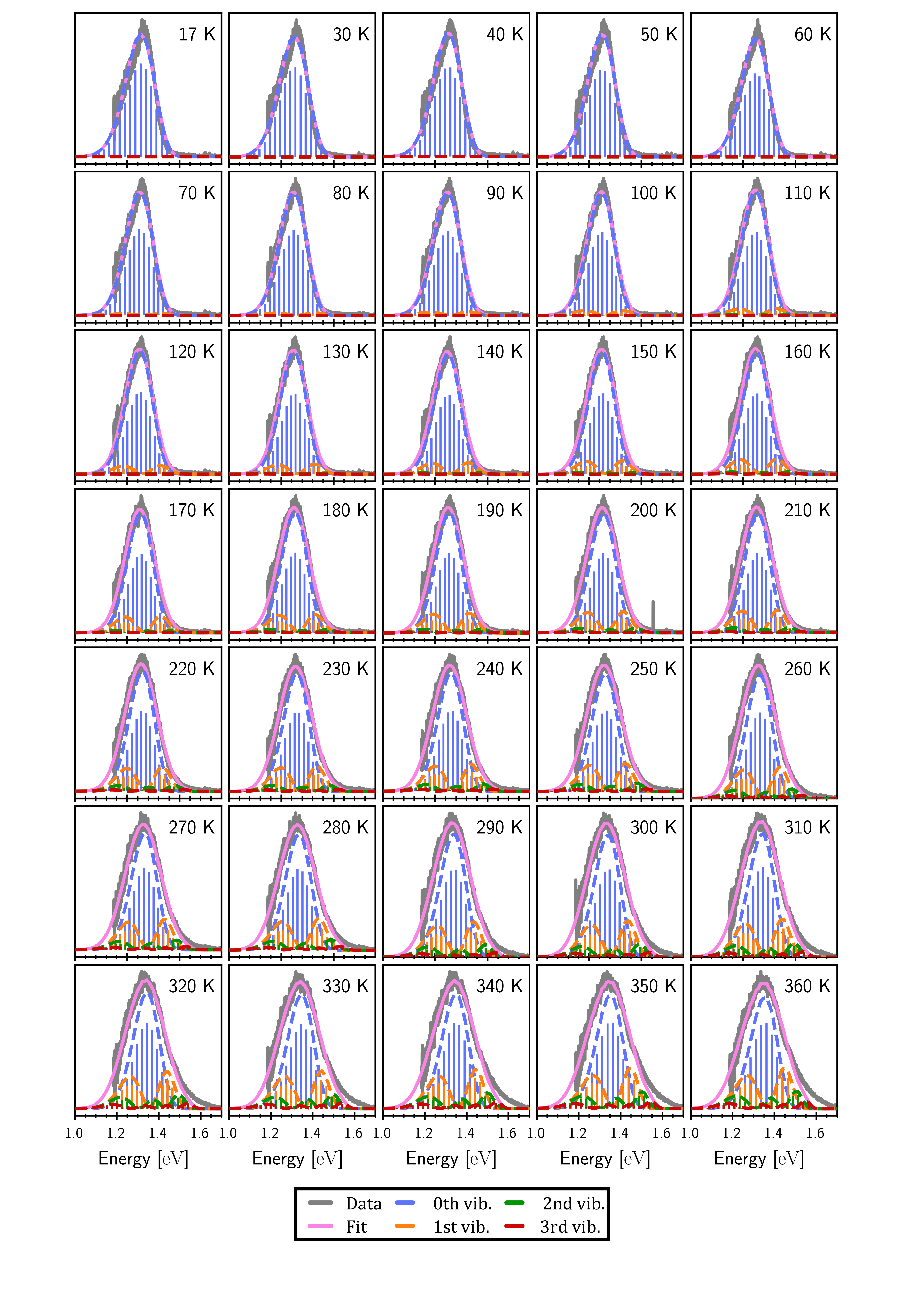}
    \caption{ZnPc emission spectra fitted with quantum mechanical X-dimer model.}
    \label{fig:S6}
\end{figure}

\begin{figure}[h]
    \centering
    \includegraphics{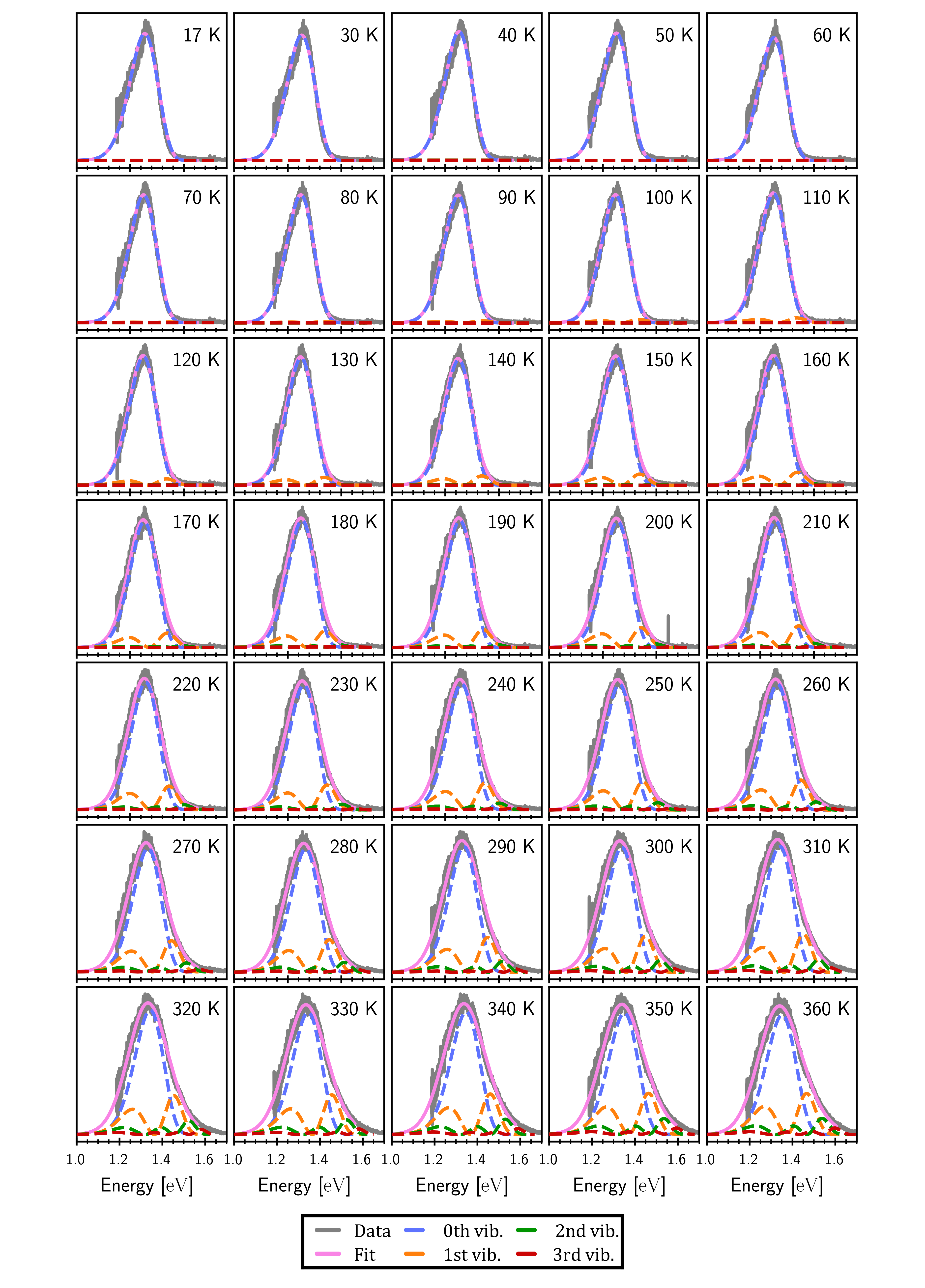}
    \caption{ZnPc emission spectra fitted with semi-classical X-dimer model.}
    \label{fig:S7}
\end{figure}

Extracted fitting parameters from the quantum mechanical and semi-classical X-dimer model are shown in figures \ref{fig:S4} and \ref{fig:S5}, respectively. Figures \ref{fig:S6} and \ref{fig:S7} show the full data set of the temperature dependent photoluminescence data including the fit curves for the quantum mechanical and semi-classical approach, respectively.

\FloatBarrier
\newpage
\FloatBarrier

\newpage
\FloatBarrier

\newpage
\printbibliography[title=Notes and references]